\documentstyle[12pt,aaspp4,flushrt]{article}
\received{RECEIPT DATE}
\revised{REVISION DATE}
\accepted{ACCEPT DATE}
\cpright{type}{year}
\journalid{VOL}{JOURNAL DATE}
\articleid{000}{000}
\paperid{MANUSCRIPT ID}

\cpright{TYPE}{YEAR}
\ccc{CODE}
\lefthead{Saglia et al.}
\righthead{Peculiar Motions of Galaxies IV}
\begin{document}
\title{The Peculiar Motions of Early-Type Galaxies in Two Distant
Regions. IV. The Photometric Fitting Procedure}
\author{R.P. Saglia}
\affil{Institut f\"ur Astronomie und Astrophysik\\
Scheinerstra\ss e 1, D-81679 Munich, Germany}
\authoraddr{Scheinerstra\ss e 1, D-81679 Munich, Germany}
\author{Edmund Bertschinger}
\affil{Dept of Physics, MIT\\Cambridge, MA 02139, U.S.A.}
\author{G. Baggley}
\affil{Dept of Astrophysics, University of Durham\\
South Road, Durham, DH1 3LE, UK}
\author{David Burstein}
\affil{Dept of Physics and Astronomy, Arizona State University\\
Tempe, AZ 85287-1504, U.S.A.}
\author{Matthew Colless}
\affil{Mount Stromlo and Siding Spring Observatories, The Australian National
University\\
Weston, Canberra ACT 2611, 
Australia}
\author{Roger L.Davies}
\affil{Dept of Astrophysics, University of Durham\\
South Road, Durham, DH1 3LE, UK}
\author{Robert K.McMahan, Jr.}
\affil{Dept of Physics and Astronomy, University of North Carolina\\
CB\#3255 Phillips Hall, Chapel Hill, NC 27599-3255, U.S.A.}
\and
\author{Gary Wegner}
\affil{Dept of Physics and Astronomy, Dartmouth College\\
Wilder Lab., Hanover, NH 03755, U.S.A.}
\newpage
\begin{abstract}

The EFAR project is a study of 736 candidate early-type galaxies in 84
clusters lying in two regions towards Hercules-Corona Borealis and
Perseus-Cetus at distances $cz\approx6000-15000$ km/s. In this paper
we describe a new method of galaxy photometry adopted to
derive the photometric parameters of the EFAR galaxies.  The algorithm
fits the circularized surface brightness profiles as the sum of two 
seeing-convolved
components, an $R^{1/4}$ and an exponential law. This approach allows us
to fit the large variety of luminosity profiles displayed
by the EFAR galaxies homogeneously  and to derive (for at least a
subset of these)
bulge and disk parameters. Multiple
exposures of the same objects are optimally combined and an optional 
 sky-fitting
procedure has been developed to correct for sky subtraction errors.
Extensive Monte Carlo simulations are analyzed to test the performance
of the algorithm and estimate the size of random and {\it systematic}
errors. Random errors are small, provided that the global
signal-to-noise ratio of the fitted profiles is larger than $\approx
300$.  Systematic errors can result from 1) errors in the sky
subtraction, 2) the limited radial extent of the fitted profiles, 3)
the lack of resolution due to seeing convolution and pixel sampling,
4) the use of circularized profiles for very flattened objects seen
edge-on and 5) a poor match of the fitting
functions to the object profiles. Large systematic errors are
generated by the widely used simple
$R^{1/4}$ law to fit luminosity profiles when a disk component, as small as
20\% of the total light, is present.

The size of the systematic errors cannot be determined from the shape
of the $\chi^2$ function near its minimum because extrapolation is
involved. Rather, we must 
estimate them by a set of quality parameters, calibrated against our
simulations, which take into account the
amount of extrapolation involved to derive the total magnitudes, the
size of the sky correction, the average surface brightness of the
galaxy relative to the sky, the radial extent of the profile, its
signal-to-noise ratio, the seeing value and the reduced $\chi^2$ of
the fit. We formulate a combined quality parameter $Q$ which indicates the
expected precision of the fits.  Errors in total magnitudes $M_{TOT}$
less than 0.05 mag and in half-luminosity radii $R_e$ less than 10\%
are expected if $Q=1$, and less than 0.15 mag and 25\% if $Q=2$; 89\%
of the EFAR galaxies have fits with $Q=1$ or $Q=2$. The errors on the
combined Fundamental Plane quantity $FP=\log R_e -0.3\langle
SB_e\rangle$, where $\langle SB_e \rangle$ is the average effective
surface brightness, are smaller than 0.03 even if $Q=3$. Thus
systematic errors on $M_{TOT}$ and $R_e$ only have  a marginal effect on the
distance estimates which involve $FP$.

We show that the sequence of $R^{1/n}$ profiles, recently used to fit
the luminosity profiles of elliptical galaxies, is equivalent (for
$n\le 8$) to a subsample of $R^{1/4}$ and exponential profiles, with
appropriate scale lengths and disk-to-bulge ratios. This suggests that
the variety of luminosity profiles shown by early-type galaxies may be due
to the presence of a disk component.
\end{abstract}


\section{Introduction}

This is the fourth paper of a series where the results of the EFAR project
are presented. In Wegner et al. (1996, hereafter Paper I) the galaxy and 
cluster sample was described, together with the related selection 
functions. Wegner et al. (1997, hereafter Paper II) reports the analysis of the
spectroscopic data. Saglia et al. (1997, hereafter Paper III) derives the 
photometric parameters of the galaxies. In this paper we describe the fitting 
technique used to derive these last quantities. 

A large number of papers have been dedicated to galaxy photometry.
The reader should refer to  the
{\it Third Reference Catalogue of Bright Galaxies} (RC3, de
Vaucouleurs et al. 1991) for a complete review of
the subject. By way of introduction we give here only a short summary of the
methods and tests adopted and performed in the past to derive the photometric
parameters of galaxies.

Using photoelectric measurements, photometric parameters
have been derived by fitting curves of growth. The RC3 values are computed
by choosing the optimal curve between a set of 15 (for $T=-5$ to
$T=10$, see Buta et al. 1995), one for each type $T$ of galaxies.
Photoelectric data are practically free from sky subtraction errors
($<0.5$\%), but can suffer from contamination by foreground objects.
Typically, 5-10 data points are available per galaxy, with apertures
which do not exceed 100 arcsec and do not always bracket the
half-luminosity diameter.  Burstein et al. (1987) (who fit the
$R^{1/4}$ curve of growth to derive the photometric parameters of a
set of ellipticals) discuss the systematic effects associated with
these procedures.  The total magnitudes $M_{TOT}$ and
effective radii $R_e$ derived are biased depending on the set of data
fitted. The errors in both quantities are strongly correlated, so that
$\Delta \log R_e -0.3 \Delta \langle SB_e\rangle\approx$ constant,
where $\langle SB_e \rangle = M_{TOT}+5\log R_e+2.5\log (2\pi)$ is the
average surface brightness inside $R_e$.
This constraint
(Michard 1979, Kormendy \& Djorgovski 1989 and references therein)
stems from the fact that the product $R{\langle I \rangle}^{0.8}$
varies only by $\pm$5\% for {\it all} reasonable growth curves (from
$R^{1/4}$ to exponential laws) in a radius range $0.5R_e\le R\le 1.5
R_e$ (see Figure 1 of Saglia, Bender \& Dressler 1993). Here $\langle
I \rangle$ is the average surface brightness inside $R$. If the
galaxies considered are large ($R_e>10$ arcsec), no seeing corrections
are needed (see Saglia et al. 1993).
 
Until the use of CCD detectors, differential luminosity profiles of galaxies
were obtained largely from photographic plates.  The procedure required
to calibrate the nonlinear response of the plates and to
digitize them is very involved. As a consequence, it was possible to
derive accurate luminosity profiles or two-dimensional photometry 
only for a small number of galaxies
(see, for example, de Vaucouleurs  \& Capaccioli 1979).
Using this sort of data, Thomsen  \& Frandsen (1983) derived $R_e$ and
$M_{TOT}$ for a set of brightest elliptical galaxies in clusters at
redshifts $<0.15$.  They fit a two-dimensional $R^{1/4}$ law convolved
with the appropriate point-spread function and briefly investigated
the systematic effects of sampling (pixel size), signal-to-noise
ratio, and shape of the profile on the derived photometric
quantities. Lauberts  \&  Valentijn (1989) digitized and calibrated the
blue and red plates of the ESO Quick Schmidt survey to derive
the photometric parameters of a large set of southern galaxies. Here
the total
magnitudes are not corrected for extrapolation to infinity, but are
defined as the integrated magnitude at the faintest measured surface
brightness (beyond the 25 B mag arcsec$^{-2}$
isophote) for which the luminosity profile is monotonically decreasing. 
In addition, the catalogue gives the parameters derived by
fitting a ``generalized de Vaucouleurs law'' 
($I=I_0\exp(-(r/\alpha)^N)$;
compare to Eq. \ref{r1n}) to the surface brightness profiles.

The last 15 years have seen the increased use of CCDs for
photometry. CCDs are linear over a large dynamic range, can be
flatfielded to better than 1\% and allow one to eliminate possible
foreground objects during the analysis of the data. Large samples of
CCD luminosity profiles for early-type galaxies have been collected by
Djorgovski (1985), Lauer (1985), Bender, D\"obereiner \& M\"ollenhoff
(1988), Peletier et al. (1990), Lucey et al. (1991), J\o rgensen,
Franx \& Kj\ae rgaard (1995).  Using CCDs one can derive photometric
parameters by fitting a curve-of-growth to the integrated surface
brightness profile.  One major concern of CCD photometry is sky
subtraction. If the CCD field is not large enough compared to the
half-luminosity radii of the galaxies, then the sky value determined
from the frame may be systematically overestimated (due to
contamination of the sky regions by galaxy light), leading to
systematically underestimated $R_e$ and $M_{TOT}$. This problem might
however be solved with the construction of very large chips or
mosaics of CCDs (see MacGillivray et al. 1993, Metzger, Luppino \&
Miyazaki 1995).

Among the most recent studies of galaxy photometry is the Medium Deep
Survey performed with the Hubble Space Telescope.  Casertano et
al. (1995) analyse 112 random fields observed with the Hubble Space
Telescope Wide Field Camera prior to refurbishment to study the
properties of faint galaxies. They construct an algorithm which fits
the two-dimensional matrix of data points to perform a disk/bulge
classification.  The $R^{1/4}$ and exponential components are
convolved with the point spread function (psf) of the HST and Monte
Carlo simulations are
performed to test the results. Disk-bulge decomposition is attempted
only for a few cases (see Windhorst et al. 1994), because the data are
in general limited by the relatively low signal-to-noise and by the spatial
resolution.

In order to derive total magnitudes, galaxy photometry involves
extrapolation of curves of growth to infinity, and therefore relies on
fits to the galaxy luminosity profiles.  Recently, Caon, Capaccioli \&
D'Onofrio (1993, hereafter CCO) and D'Onofrio, Capaccioli \& Caon
(1994) focused on the use of the $R^{1/4}$ law to fit the photometry
of ellipticals.
%Burkert (1993) finds systematic deviations which
%correlate with the galaxy isophotal shapes (disky versus boxy, see
%Bender et al. 1988). 
CCO find a correlation with the galaxy size and argue that if an
$R^{1/n}$ law (Sersic 1968, see Eq. \ref{r1n}) is used to fit the
luminosity profiles, then smaller galaxies ($\log R_e[{\rm kpc}]<0.5$) are best
fitted with exponents $1<n<4$, while larger ones ($\log R_e[{\rm
kpc}]>0.5$) have $n>4$. Half-light radii and total magnitudes derived
using these results may differ strongly from those using $R^{1/4}$
extrapolations. Finally, Graham et al. (1996) find that the extended
shallow luminosity profiles of BCG are best fit by $R^{1/n}$ profiles 
with $n>4$.

To summarize, 
the EFAR collaboration has collected photoelectric and CCD photometry
for 736 galaxies (see Colless et al. 1993 and Paper III), 31\% of
which appear to be spirals or barred objects. The remaining
69\% can be subdivided in cD-like (8\%), pure E (12 \%) and mixed E/S0
(49\%); the precise meaning of these classifications is explained in
detail in \S 3.4 of Paper III.  
We derived circularly averaged luminosity profiles for all of
the objects. Isophote shape analysis can only be reliably performed for the
subset of our objects which are large and bright enough, and will be 
discussed in
a future paper. Since 96\% of the EFAR galaxies have ellipticities
smaller than 0.4, the use of circularized profiles does not introduce
systematic errors on the photometric parameters derived (see \S
\ref{decomposition}) and has the advantage of giving robust results
for even the smaller, fainter objects in our sample.  
The galaxies show a large variety of profile
shapes. Typically, each object has been  observed several
times, using a range of telescopes, CCD detectors, and exposure
times, under different atmospheric and seeing conditions, with
different sky surface brightnesses.  

Deriving homogeneous photometric parameters from the large EFAR data
set has required the construction of a sophisticated algorithm to (i)
optimally combine the multiple photoelectric and CCD data of each
object, (ii) fit the resulting luminosity profiles with a model
flexible enough to describe the observed variety of profiles, (iii)
classify the galaxies morphologically and (iv) produce reliable
magnitudes and half-luminosity radii.

This paper describes our method as applied in Paper III. It explores
the sources of random and systematic errors by means of Monte Carlo
simulations, and develops a scheme to quantify the precision of the
derived parameters objectively. The fitting algorithm searches for the
best combination of the seeing-convolved, sky-corrected $R^{1/4}$ and
exponential laws. This approach fulfils the requirement (ii) above: it
produces convenient fits to the extended components of cD luminosity
profiles, it models the profile range observed in E/S0 galaxies (from
galaxies with flat cores to clearly disk-dominated S0s), and it
reproduces the surface brightness profiles of spirals. Moreover, for
the E/S0s and spirals, this approach determines the parameters of
their bulge and disk components, to assist classification (requirement
(iii)). Finally, this approach minimizes extrapolation (requirement
(iv)) which is the main source of uncertainty involved in the
determination of magnitudes and half-luminosity radii.

Would it be possible to reach the same goals with another choice of
fitting functions? We demonstrate here (\S \ref{profiles}) that the
$R^{1/n}$ profiles quoted above can be seen as a ``subset'' of the
$R^{1/4}$ plus exponential models and therefore might not meet
requirement (ii). In addition, for $n>4$ they require large
extrapolations and therefore might fail to meet requirement (iv). What
is the physical interpretation of the two components of our fitting
function? There are cases (the above cited cD galaxies and the
galaxies with cores) where our two-component approach provides a good
fitting function, but the ``disk-bulge'' decomposition is not
physical. However, we argue that the systematic deviations from a
simple $R^{1/4}$ law observed in the luminosity profiles of our
early-type galaxies are the signature of a disk. We will investigate
this question further in a future paper, where the isophote shape
analysis of the largest and brightest galaxies in the sample will be
presented. Would it be worth improving the present scheme by, for
example, allowing for a {\it third} component (a second $R^{1/4}$ or
exponential) to be fit? This could produce better fits to barred
galaxies or to galaxies with cores and extended shallow
profiles. However, it is not clear that the systematic errors related
to extrapolation and sky subtraction could be reduced. Summarizing,
the solution adopted here fulfils our requirements (i)-(iv).

This paper is organized as follows.  \S \ref{fitting} describes the
three-step fitting technique. This involves the algorithm for the
combination of multiple profiles of the same object (\S
\ref{combination}), our two-component fitting technique with the
additional option of sky fitting (\S \ref{diskbulge}), and the
objective quality assessment of the derived parameters (\S
\ref{quality}). \S \ref{montecarlo} presents the results of the Monte
Carlo simulations performed to test the fitting
procedure and assess the precision of the derived photometric
parameters. We explore a large region of the parameter space ($R_{eB},
h$, D/B, $\Gamma$, see \S \ref{diskbulge} for a definition of the
parameters) and test the performance of the fitting algorithm (\S
\ref{parameter}).  In \S \ref{sky} we investigate the systematic
effects introduced by possible errors on sky subtraction and test the
algorithm to correct for this effect (see \S \ref{diskbulge}).  The
influence of the limited radial extent of the profiles (\S
\ref{extension}), of the signal-to-noise ratio (\S \ref{snratio}), and
of seeing and pixellation (\S \ref{seeing}) are also investigated. The
profile combination algorithm is tested in \S \ref{testcombination}.
In \S \ref{decomposition} we assess the effectiveness of using the
fitting algorithm to derive the parameters of bulge and disk
components of a simulated galaxy.  A number of different 
profiles are considered in \S \ref{profiles} to test their systematic
effect on the photometric parameters. We show that the $R^{1/n}$
profiles can be reproduced by a sequence of $R^{1/4}$ plus exponential
profiles, with small systematic differences ($<0.2$ mag arcsec$^{-2}$)
over the radial range $R_e/20<R<5R_e$ (see discussion above). 
In \S \ref{discussion} we discuss
how to estimate the precision of the derived photometric parameters.
In \S \ref{conclusions} we summarize our results in terms of the
expected uncertainties on the derived photometric parameters.
%\newpage

\section{The fitting procedure}
\label{fitting}

The algorithm devised to fit the luminosity profiles of EFAR galaxies
(see Paper III) involves three connected steps, (i) the combination of
multiple profiles, (ii) the two-component fitting, and (iii) the
quality estimate of the results. In the first step, the multiple CCD
luminosity profiles available for each object are combined taking into account
differences in sensitivity or exposure time, and sky subtraction
errors.  A set of multiplicative and additive constants is determined
($k_i$, $\Delta_i$), which describe respectively the relative scaling
due to sensitivity and exposure time and the relative difference in
sky subtraction errors. The absolute value of the scaling is the
absolute photometric calibration of the images. This is accomplished
as described in Paper III, making use of the photoelectric aperture
magnitudes and absolute CCD calibrations. The absolute value of the
sky correction can be fixed either to zero or to a percentage of the
mean sky, or passed to the second step to be determined as a result of the
fitting scheme.

The second step fits these combined profiles. The backbone of the
fitting algorithm is the sum of the seeing-convolved $R^{1/4}$ and the
exponential laws.  We have discussed the advantages of this choice in
the Introduction. This combination produces a variety of luminosity
profiles which can fit a large number of realistic profiles to high
accuracy. The photometric parameters derived from this approach do not
require large extrapolations, if the available profiles extend to at
least $4R_e$. When galaxies with disk and bulge components (E/S0s and
spirals) are seen at moderate inclination angles (as it is the case
for the EFAR sample, where 96\% of galaxies have ellipticities less
than 0.4, see Paper III), then the algorithm is also able, to some
extent, to determine the parameters of the two components. In Paper
III this information is used, together with the visual inspection of
the images and, sometimes, the spectroscopic data, to classify each
EFAR object as E, E/S0, or spiral. While we believe that in these
cases the two components of the fits are indicative of the presence
of two physical components, additional investigation is certainly required to
test this conclusion. This will involve the 
isophote shapes analysis (Scorza \& Bender 1995), the fitting of the
two-dimensional photometry (Byun \& Freeman 1995, de Jong 1996), the
colors and metallicities (Bender \& Paquet 1995), and the kinematics
(Bender, Saglia \& Gerhard 1994) of the objects. We intend to address 
some of these issues in future papers for a selection of large and
bright EFAR galaxies.

The third step assigns quality parameters to the derived photometric 
parameters. Several factors determine how accurate these parameters can 
be expected to be. \S \ref{montecarlo} explores in detail the effects of 
sky subtraction errors,   radial extent, signal-to-noise ratio, 
seeing and sampling, and goodness of fit. A global quality parameter 
based on these  results quantifies the precision of the final results. 

%\newpage
\subsection{Profile combination}
\label{combination}

The first step of the fitting algorithm is to combine the multiple
profiles available for each galaxy. Fitting each profile separately,
and averaging the
results produces severely biased
results if the fitted profiles differ in their signal-to-noise ratio,
seeing and sampling, radial extent, and sky subtraction errors. Only
a simultaneous fit can minimize the biasing effects of
these factors (see \S \ref{testcombination}).

Apart from the very central regions of galaxies, where seeing and
pixel size effects can be important, the profiles of the same object
taken with different telescopes and instruments differ by 
a normalization (or multiplicative constant) only and an additive
constant. The first takes into account differences in the 
efficiency  and transparency, while the second adjusts for the relative
errors in the sky subtraction.  Let $I_i(R)$, $i=1$ to $n$ denote the
$n$ available
profiles in counts per arcsec$^2$ at a distance $R$ from the center,
and consider the profile $I_{max}(R)$ as the one having the maximum radial
extent. In general the radial grids on which the profiles $I_i(R)$
have been measured will not be the same, but it will always be
possible to (spline) interpolate the values of $I_{max}(R)$ on each of
the grid points of the other profiles $I_i(R)$.  The normalization $k_i$ of 
the 
profiles $I_i(R)$ relative to the profile $I_{max}(R)$ and the quantity
$\Delta_i$ (related to $\Delta_i/k_i$  the correction to the sky value
of the profile $I_i(R)$)  are the
multiplicative and additive constants to be sought, so that:
\begin{equation}
\label{corrected}
I_i'(R)=k_iI_i(R)-\Delta_i. 
\end{equation}
The $k_i$ and $\Delta_i$ constants are determined by minimizing the
$\chi^2$-like functions (see the related discussion for
Eq. \ref{chitot}):
\begin{equation}
\label{chid}
\chi^2_i=\sum_{R>R_c} w_i(R)\left(I_{max}(R)-k_i I_i(R)+\Delta_i\right)^2.
\end{equation}
The inner cutoff radius $R_c$ is 6 arcsec or half of the maximum
extent of the profile, if this is less than 6 arcsec. This cutoff 
minimizes the influence of seeing, while retaining a
reasonable number of points in the sums.  Here
$w_i(R)=1/\sigma_i(R)^2$ are the relative weights of the data points,
which are related to the expected errors for the profile
$I_i$:
\begin{equation}
\label{weightw}
\sigma_i(R)=\frac {\sqrt{G_i I_i(R)+G_i\hbox{Sky}_i+RON_i^2/S_i^2}}
{\sqrt{2\pi R/S_i}},
\end{equation}
where $S_i$, $G_i$ and $RON_i$ are the scale (in arcsec/pixel), the gain and
the readout noise of the CCD used to obtain the profile $I_i$ (see
Table 2 of Paper III). The denominator of Eq. \ref{weightw} assumes
that all of the pixels in the annulus at radius $R\neq 0$ have been averaged 
to get $I(R)$ and therefore underestimates the 
errors if some pixels have been masked to delete background or 
foreground objects superimposed on the program galaxies. If $R=0$ (i.e., the
central pixel) the following equation is used:
\begin{equation}
\label{weightw0}
\sigma_i(R=0)={\sqrt{G_i I_i(R=0)+G_i\hbox{Sky}_i+RON_i^2/S_i^2}}.
\end{equation}
The weight in this fit monotonically increases with radius.
The errors $\sigma_{\mu_i}$  on the surface brightness magnitudes 
$\mu_i=-2.5 \log I_i$ are related to Eqs. \ref{weightw}  and
\ref{weightw0} through:
\begin{equation}
\label{sigmai}
\sigma_{\mu_i}=\frac{2.5  \sigma_i(R)\log e}{I_i(R)},
\end{equation}
By requiring $\partial \chi^2_i/ \partial k_i=0$ and $\partial 
\chi^2_i/ \partial \Delta_i=0$ we solve the linear system
for $k_i$ and $\Delta_i$.

%
%\begin{equation}
%\label{eqk}
%k_i=\frac{\sum_{R>R_c} w_i(R)I_{max}(R)I_i(R)\sum_{R>R_c} v_i(R)-\sum_{R>R_c} v_i(R)I_i(R)\sum_{R>R_c}
% w_i(R)I_{max}(R)}
%     {\sum_{R>R_c} w_i(R)I_{max}^2(R)\sum_{R>R_c} v_i(R)-\sum_{R>R_c} v_i(R)I_{max}(R)\sum_{R>R_c} w_i(R)I_{max}(R)},
%\end{equation}
%
%\begin{equation}
%\label{eqd}
%\Delta_i=-\frac{\sum_{R>R_c} w_i(R)I_{max}(R)I_i(R)\sum_{R>R_c} v_i(R)I_i(R)-\sum_{R>R_c} v_i(R)I_{max}(R)\sum_{R>R_c} w_i(R)I_i^2(R)}
%     {\sum_{R>R_c} w_i(R)I_{max}^2(R)\sum_{R>R_c} v_i(R)-\sum_{R>R_c} v_i(R)I_{max}(R)\sum_{R>R_c} w_i(R)I_{max}(R)}.
%\end{equation}

At this stage the relative sky corrections are known for all of the
profiles except the most extended one. This last correction
$\Delta_{max}$ can either be computed as part of the fitting program
(see Eqs. \ref{chikmax} and \ref{chidmax}), or fixed to a given value.

In \S \ref{montecarlo} the strategy of setting the mean 
percentage sky errors (for a given galaxy) to zero will be tested
extensively against the above. For this case one requires:
\begin{equation}
\label{meanzero}
\frac{\Delta_{max}}{\hbox{Sky}_{max}}+\sum_i \frac{\Delta_i}{k_i \hbox{Sky}_i}=0.
\end{equation}
In general, Eq. \ref{meanzero} is not a good choice and gives rise to
systematic errors (see Fig. \ref{figskyerror}), however it is 
preferred when the sky fitting solution (Eq. \ref{chidmax}) requires 
excessively  large extrapolations. Forty percent
of the fits presented in Paper III 
are performed using Eq. \ref{meanzero}.

Note that for both Eq.  \ref{meanzero} and \ref{chidmax} described
below, the
value of $\Delta_{max}$ is determined iteratively, 
by minimizing Eq. \ref{chid}, having
redefined $I_{max}(R)$ as $I'_{max}(R)$, where
$I'_{max}(R)=I_{max}(R)-\Delta_{max}$, and repeating the procedure
until it convergences.  Four or five iterations are needed to
reach a precision $<10^{-5}$ when Eq. \ref{meanzero} is used. 
Convergence is reached while performing the non-linear fitting of \S
\ref{diskbulge}, when using Eq. \ref{chidmax}. Sky corrections, as computed
in Paper III, are less than 1 \% for 80\% of the cases examined.

The absolute scaling, $k_{max}$, of the $I_{max}(R)$ profile represents the
photometric calibration of the profiles. This is performed as described 
in Paper III using the photoelectric aperture magnitudes and CCD zero-points. 
In the following we set $k_{max}=1$.

\subsection{$R^{1/4}+$exponential law fitting}
\label{diskbulge}

The surface brightness profiles of each galaxy are 
modeled simultaneously  by assuming that they can be represented by
the sum of a de Vaucouleurs law (the ``bulge'' component indicated by B) 
and an exponential component (the ``disk'' component indicated by D):
\begin{equation}
\label{fittingfun}
f(R,R_{eB},h,D/B,\Gamma,S)_{B+D}=f_B+f_D,
\end{equation}
where $R_{eB}$ is the half-luminosity radius of the bulge component,
$h$ the exponential scale length of the disk component, $D/B$ the disk
to bulge ratio, $\Gamma$ the FWHM of the seeing profile, and $S$ the
pixel size.  Both laws are seeing-convolved as described by Saglia et
al. (1993) and take into account the effects of finite pixel
size. Definitions and numerical details can be found in the Appendix.
The results presented in Paper III  show
that Eq.  \ref{fittingfun} gives fits with respectably small
residuals. The differences in surface brightness $\Delta
\mu=\mu-\mu_{fit}$ are typically less than 0.05 mag arcsec$^{-2}$, while those
between the integrated aperture magnitudes are a factor two
smaller. However our formal values of reduced $\chi^2$ (see discussion
below) indicate that very few galaxies (less than 10\%) have
luminosity profiles that are fit well by the model disk and bulge. Over 90\%
of the fits have reduced $\chi^2$ larger than 2. In this sense
Eq. \ref{fittingfun} is not a statistically good representation of the
galaxy profiles.

A hybrid non-linear least squares algorithm is used to find the
$R_{eB}$, $h$, D/B and the vector of seeing values which gives the
best representation $f_{B+D}(R)$ of the profiles $I_i(R)$, taking into
account the sky corrections $\Delta_i/k_i$. The algorithm uses the
Levenberg-Marquardt search (Press et al. 1986), repeated several times
starting from randomly scattered initial values of the parameters. The
search is repeated using the Simplex algorithm (Press et
al. 1986). The best of the two solutions found is finally chosen.
This approach minimizes the biasing influence of the possible presence
of several nearly-equivalent minima of Eq. \ref{chitot}, a problem
present especially when low disk-to-bulge ratios are considered (see
discussion in \S \ref{parameter}).

All of the 
profiles $I_i(R)$ available for a given galaxy are fitted simultaneously 
determining  the appropriate value of the 
seeing $\Gamma_i$, for each single profile $i$.
The minimization is performed on the function:
\begin{equation}
\label{chitot}
\chi^2_{totB+D}=\sum_i\left(\sum_{R,\lambda_i f_{B+D}>-\Delta_i/k_i} 
T_>^2+\sum_{R,\lambda_i f_{B+D}<-\Delta_i/k_i} T_<^2\right),
\end{equation}
where:
\begin{equation}
\label{termmaj}
T_>=-2.5\log \left[\frac{\lambda_i f_{B+D}(R,R_{eB},h,D/B,\Gamma_i,S_i)+\Delta_i/k_i}{I_i(R)}\right]\frac{p_i}{\sigma_{\mu_i}},
\end{equation}
and:
\begin{equation}
\label{termmin}
T_<=-2.5 \log \left[\frac{\lambda_i f_{B+D}(R,R_{eB},h,D/B,\Gamma_i,S_i)}
{I_i(R)-\Delta_i/k_i}\right]\frac{p_i}{\sigma_{\mu_i}}
\end{equation}

The penalty function $p_i$ is introduced to avoid unphysical solutions
and increases $\chi^2_{totB+D}$ to very large values when $D/B<0$ or
when the values of of $R_{eB}$ or $h$ become too large ($>300''$) or
too small ($<1''$). The use of the
$T_>$ and $T_<$ terms ensures that the arguments of the logarithm are
always positive. The sky correction is usually applied to the fitting
function (see Eq. \ref{termmaj}). However, data points where
$\lambda_i f_{B+D}+ \Delta_i/k_i<0$ (this may happen when a negative
sky correction $\Delta_i/k_i$ is applied) are included using
Eq. \ref{termmin}, which applies the sky correction to the data
points.  Note also that Eq. \ref{chitot} is the weighted sum of the
squared {\it magnitude} residuals. This is to be preferred to the
weighted sum of the squared linear residuals, which is dominated by the data
points of the central parts of the galaxies.

The model normalization  relative to the profile $I_i(R)$, $\lambda_i$,
is determined by requiring
$\partial \chi^2_{\lambda_i}/ \partial \lambda_i=0$, where:
\begin{equation}
\label{chikmax}
\chi^2_{\lambda_i}=\sum_R w_i(R)\left(I_i(R)-\lambda_i f_{B+D}(R)- 
\Delta_i/k_i\right)^2.
\end{equation}
Note that the ratios $\lambda_{max}/\lambda_i$ can in principle differ
from the constants $k_i$, because of (residual) seeing effects (see,
e.g., $R_c$ in Eq. \ref{chid}) and systematic differences between
model and fitted profiles. In fact, the differences are smaller than
8\% in 85\% of the fits performed with more than one profile (see
Paper III).  When a bulge-only or a two-component model is used, the
total magnitude of the fitted galaxy, in units of the $I_{max}(R)$
profile, is computed as $M_{TOT}=-2.5 \log (L_B+L_D)$, where
$L_B=\lambda_{max} R_{eB}^2$ (see Eq. \ref{bulge}, with this
normalization one has $I_{eB}=\lambda_{max}/(7.22\pi)$) is the
luminosity of the bulge and $L_D=(D/B)L_B$ is the luminosity of the
disk. When a disk-only model is used, then $M_{TOT}=-2.5 \log L_D$,
where $L_D=\lambda_{max}h^2$ (see Eq. \ref{disk}, with this
normalization one has $I_0=\lambda_{max}/(2\pi)$). 
Note again that the photometric
calibration of these magnitudes $M_{TOT}$ to apparent magnitudes $m_{T}$ is
performed in Paper III using photoelectric aperture magnitudes and CCD
zero-points.

The sky correction to the profile $I_{max}$ can be set to a given
value (zero for no sky correction, using Eq. \ref{meanzero} for zero
mean percentage sky correction). Alternately, a fitted sky
correction $\Delta_{max}$ can be determined by additionally requiring
$\partial \chi^2_{\lambda_{max}}/ \partial \Delta_{max}=0$, where:
\begin{equation}
\label{chidmax}
\chi^2_{\lambda_{max}}=\sum_R w_{max}(R)\left(I_{max}(R)-\lambda_{max}f_{B+D}(R)-
\Delta_{max}\right)^2.
\end{equation}
If the resulting $\Delta_{max}$ produces
$\lambda_{max}f_{B+D}+\Delta_{max}<0$ at any $R$,  
Eq. \ref{termmin} is used to compute 
the corresponding contribution to Eq. \ref{chitot}.  When using
Eq. \ref{chidmax}, the constants $k_i$ and $\Delta_i$ are computed
again using $I'_{max}(R)=I_{max}(R)-\Delta_{max}$ (see the profile
combination iterative algorithm in \S \ref{combination}).  The
Monte Carlo simulations of \S \ref{montecarlo} show that
Eq. \ref{chidmax} gives an unbiased estimate of the sky corrections
when the $f_{B+D}$ is a good model of the fitted profiles. 
Eq. \ref{meanzero} is to be preferred when large extrapolations are
obtained; 60\% of the
fits presented in Paper III  are performed using
Eq. \ref{chidmax}. 

One might use the equivalent of Eq.
\ref{chidmax} for the profiles $I_i(R)$ to compute the corrections
$\Delta_i$ directly from the fit, without having to go through Eq. \ref{chid}.
This would automatically take into account the seeing differences 
of the profiles. However, tests show that this approach does not produce
the correct relative sky corrections between the profiles, if the fitting 
function does not describe the fitted profiles well.
Finally, one might try deriving $\lambda_i$ and $\Delta_i$ by minimizing
Eq. \ref{chitot} for these two additional parameters. The adopted
solution, however,
speeds up the CPU intensive, non-linear minimum search, since 
$\lambda_i$ and $\Delta_i$ are computed analytically. 

The fit is repeated using a pure de Vaucouleurs law (D/B=0) and a pure
exponential law (B/D=0).  In analogy with Eq. \ref{chitot}, two other
$\chi^2_{tot}$ are considered for these fits, $\chi^2_{totB}$ and
$\chi^2_{totD}$. A (conservative) $3\sigma$ significance test (see discussion
after Eq. \ref{chido}) is performed to decide whether
the addition of the second component  improves the fit significantly.
The bulge-only fit is taken if:
\begin{equation}
\label{chibo}
\frac{\chi^2_{totB}}{\chi^2_{totB+D}}-1<3\sqrt{\frac{2}{N^{free}_{B+D}}}.
\end{equation}
The disk-only fit is taken if:
\begin{equation}
\label{chido}
\frac{\chi^2_{totD}}{\chi^2_{totB+D}}-1<3\sqrt{\frac{2}{N^{free}_{B+D}}}.
\end{equation}

The number of degrees of freedom of the $R^{1/4}$ plus exponential law
fit is $N^{free}_{B+D}=N_{data}-N_{sky}-3-2N_{prof}$, where $N_{data}$ is
the number of data points involved in the sum of Eq. \ref{chitot},
$N_{sky}=1$ if the sky fitting is activated, zero otherwise, and
$3+2N_{prof}$ are the number of parameters fitted ($R_{eB},h,D/B$,
$M_{TOT}$, $N_{prof}$ seeing values and $N_{prof}-1$ normalization constants
$\lambda_i$, where $N_{prof}$ is the number of fitted profiles).  If
the errors $\sigma_{\mu_i}$ are gaussian, $\chi^2_{totB+D}$
follows a $\chi^2$ distribution of $N^{free}_{B+D}$ degrees of freedom.
If the bulge plus disk model is a good representation of the data,
then the $\chi^2_{totB+D}\approx N^{free}_{B+D}$ in the mean, with an
expected dispersion $\sqrt{2N^{free}_{B+D}}$. In this case
Eqs. \ref{chibo} and \ref{chido} are a 3$\sigma$ significance test on
the conservative side, meaning that one-component models are
preferred, if two-component models do not improve the fit by more than
$3\sigma$. In fact, Paper III shows that only 10\% of the
fits are statistically ``good'' ($\chi^2_{totB+D}\approx
N^{free}_{B+D}$). The median reduced $\chi^2$,
$\hat\chi^2=\chi^2_{totB+D}/ N^{free}_{B+D}$, is $\approx 6$, indicating
the existence of statistically significant systematic deviations from
the simple two-component models of Eq. \ref{fittingfun}. Fortunately,
the tests performed in \S \ref{montecarlo} show that reliable
photometric parameters can be obtained even in these cases. Note that
fits based on the $R^{1/n}$ profiles {\it do not} give better results:
Graham et al. (1996) obtain reduced $\chi^2\approx 10$ for their
sample of brightest cluster galaxies. Eqs. \ref{chibo}
and Eq.  \ref{chido} as applied in Paper III select a bulge-only fit in
14 \% of the cases, and a disk-only fit in less than 1\%. In the 85\%
of the cases when both components are used, the median value of the
significant test is 16$\sigma$, with significance larger than
5$\sigma$ in 90\% of the cases. In the following
sections and plots we shall indicate the reduced $\chi^2$ with $\chi^2$.

Total magnitudes of galaxies are extrapolated values.  In order to
quantify the effect of the extrapolation, we also derive the percentage
contribution to $M_{TOT}=-2.5 \log (L_B+L_D)$ due to the extrapolated
light beyond the radius $R_{max}$ of the last data point. In 80\%
of the galaxies examined in Paper III this extrapolation is less than 10\%.
 The half-luminosity radius $R_e$ (and the $D_n$ diameter, see Paper III) of
the best-fitting function is computed using Eqs. \ref{bulgegrowth} and
\ref{diskgrowth}, so that seeing effects are taken into account.
Finally, the contamination of the sky due to galaxy light is estimated
by computing the mean surface brightness in the annulus with radii
$R_i^{max}$ and $2R_i^{max}$, where $R_i^{max}$ is the radius of the
last data point of the profile $i$. Galaxy light contamination is less
than 0.5 \% of the sky in 80\% of the cases studied in Paper III. 

Using the appropriate seeing-convolved tables (see \S \ref{diskbulge}), 
the fitting algorithm can also be used to fit a $f_\infty$ $\Psi=12$ model 
(see description in Saglia et al. 1993 and the appendix here) plus
exponential, or a smoothed $R^{1/4}$ law plus exponential. These
additional fitting models are useful to study the effects of the
central concentration and radial extent of galaxies (see \S \ref{seeing}).

\subsection{Quality parameters}
\label{quality}

The third step in the fitting procedure assigns quality estimates to
the derived photometric parameters. Several factors determine their
expected accuracy. (i) Low signal-to-noise
images provide fits with large random errors. (ii) Images
of small galaxies observed under poor seeing conditions and/or with
inadequate sampling (a detector with large pixel size) give systematically
biased fits. (iii) Images of large galaxies taken with a small detector
give profiles with too little radial extent and fits involving
large, uncertain extrapolations.  (iv) Sky subtraction errors bias the
faint end of the luminosity profiles and therefore the fitted
parameters. Finally, (v) bad fits to the luminosity profiles provide biased
quantities. The effects of these possible sources of errors
are estimated by means of Monte Carlo simulations in \S \ref{montecarlo}.

Based on these results, one can assign the quality estimates
$Q_{max}$, $Q_\Gamma$, $Q_{S/N}$, $Q_{\hbox{Sky}}$, $Q_{\delta \hbox{Sky}}$, $Q_{E}$,
$Q_{\chi^2}$ according to the rules listed in Table \ref{tabquality},
where increasing values of the quality estimates correspond to
decreasing expected precision of the photometric parameters derived
from the fits. The global quality parameter $Q$:
\begin{equation}
\label{qtot}
Q=Max(Q_{max},Q_{\Gamma},Q_{S/N},Q_{\hbox{Sky}},Q_{\delta \hbox{Sky}},Q_E,Q_{\chi^2}),
\end{equation}
assumes values $1,2,3$, corresponding to expected precisions on total 
magnitudes $\Delta M_{TOT}
\approx 0.05,0.15,0.4$, on the logarithm of the half-luminosity radius 
$\Delta \log R_e \approx 0.04, 0.1,0.3$ and on the combined quantity 
$FP=\log R_e-0.3\langle SB_e \rangle$ $\Delta FP
\approx 0.005, 0.01,0.03$ (see \S \ref{discussion}, Fig. \ref{figqualdis}). 
Paper III shows that 16\% of EFAR galaxies have
$Q=1$, 73\% have $Q=2$ and 11\% $Q=3$.

Note that $M_{TOT}$ and, therefore, $FP$ are subject to the additional 
uncertainty due to the photometric zero-point. In Paper III we extensively
discuss this source of error and find that it is smaller than 0.03 mag
per object, for all of the cases (86\%) where a photoelectric or a CCD
calibration has been collected.

%\placetable{tabquality}
\begin{table}[ph]
\caption[*]{The definition of the quality parameters} 
\label{tabquality}
\begin{tabular}{cccccccc}
&&&&&&&\\
\tableline
\tableline
$R_{max}/R_e^f$ & $Q_{max}$ & $R_e^f/\Gamma^f$ & $Q_\Gamma$ & $S/N$ & $Q_{S/N}$ &
Extrap & $Q_E$ \\
\tableline
$\le 1$     & 3  & $\le 2$ & 2 & $\le 300$ & 2 & $\ge0.3$ & 3 \\
$>1$,$\le2$ & 2  & $>2$    & 1 & $>300$    & 1 & $<0.3$ & 1 \\
$>2$        & 1  &         &   &           &   &        &   \\ \hline
&&&&&&&\\
\tableline
\tableline
$\chi^2$          & $Q_{\chi^2}$ & $\mu_{\hbox{Sky}}-\langle SB_e^f \rangle$ & $Q_{\hbox{Sky}}$ & 
$|\delta \hbox{Sky/Sky}|$ & $Q_{\delta \hbox{Sky}}$ & & \\
\tableline
&&&&&&&\\
 $\ge25$          & 3  & $\le0.75$ & 2 & $>0.03$         & 3 & &\\ 
 $\ge12.5$, $<25$ & 2  & $>0.75$ & 1 & $>0.01$,$<0.03$ & 2 & &\\
 $<12.5$          & 1  &         &   & $<0.01$         & 1 & &\\
\tableline
\end{tabular}
\end{table}

%\newpage
\section{Monte Carlo Simulations}
\label{montecarlo}

The fitting procedure described in the previous section has been
extensively tested on simulated profiles with the goals of
checking the minimization algorithm and quantifying the effects of the
errors described in \S \ref{quality}. Luminosity
profiles of models with known parameters have been fitted, to compare
input and output values. In all of the following figures the output
parameters of the fit are indicated with the superscript $f$ for
``fit'' (for example, $\Gamma^f$).

As a first step (\S \ref{parameter}-\ref{testcombination}, Figures
\ref{figresdb}-\ref{figcombination}), we ignore possible systematic
differences between test profiles and fitting functions (such as the
ones possibly present when fitting real galaxies, see discussion in
Paper III) and generate a number of $R^{1/4}$ plus exponential model
profiles of specified $R_{eB}$, $h$, $D/B$ ratio, seeing $\Gamma$ and
total magnitude, using the seeing-convolved tables described in the
Appendix.  A constant can be added (subtracted) to simulate an
underestimated (overestimated) sky subtraction.  Given the pixel size,
the sky value, the gain and readout noise, appropriate gaussian noise
is added to the model profile following Eqs. \ref{weightw} and
\ref{weightw0}. The maximum
extent of the profiles can be specified to simulate the finite size of
the CCD. The profile is truncated at the radius where noise (or the
sky subtraction error) generates negative counts for the model.
The signal-to-noise ratios computed in the
following refer to the total 
number of counts in the model profile out to this radius.  The
parameter space explored in all of the simulations discussed in \S
\ref{parameter}-\ref{seeing} is displayed in Figure \ref{figparspace}
and covers the region where the EFAR galaxies are
expected to reside (see Paper III). Different symbols identify the
models (see caption of Fig.  \ref{figparspace}).  As a second step (\S
\ref{decomposition}-\ref{profiles}, Figures
\ref{figbulgedisk}-\ref{figr1npar}), we explore the influence of
systematic differences between test profiles and fitting functions. 
In \S \ref{decomposition} we show that fitting circularized profiles
of moderately flattened galaxies (as the ones observed in Paper III)
allows good determinations of the photometric parameters and also of
the bulge and disk components. In \S \ref{profiles} we fit the
$R^{1/n}$ profiles, achieving two results. First, we quantify the
influence of the quoted systematic effects on the fitted photometric
parameters. Second, we suggest that the possible correlation between
galaxy sizes and exponent $n$ (see discussion in the
Introduction) reflects the presence of a disk component
in early-type galaxies.  
\S \ref{discussion} summarizes the results by calibrating the quality
parameter $Q$ of Eq. \ref{qtot}.

\subsection{The parameter space}
\label{parameter}

\begin{figure}
\plotone{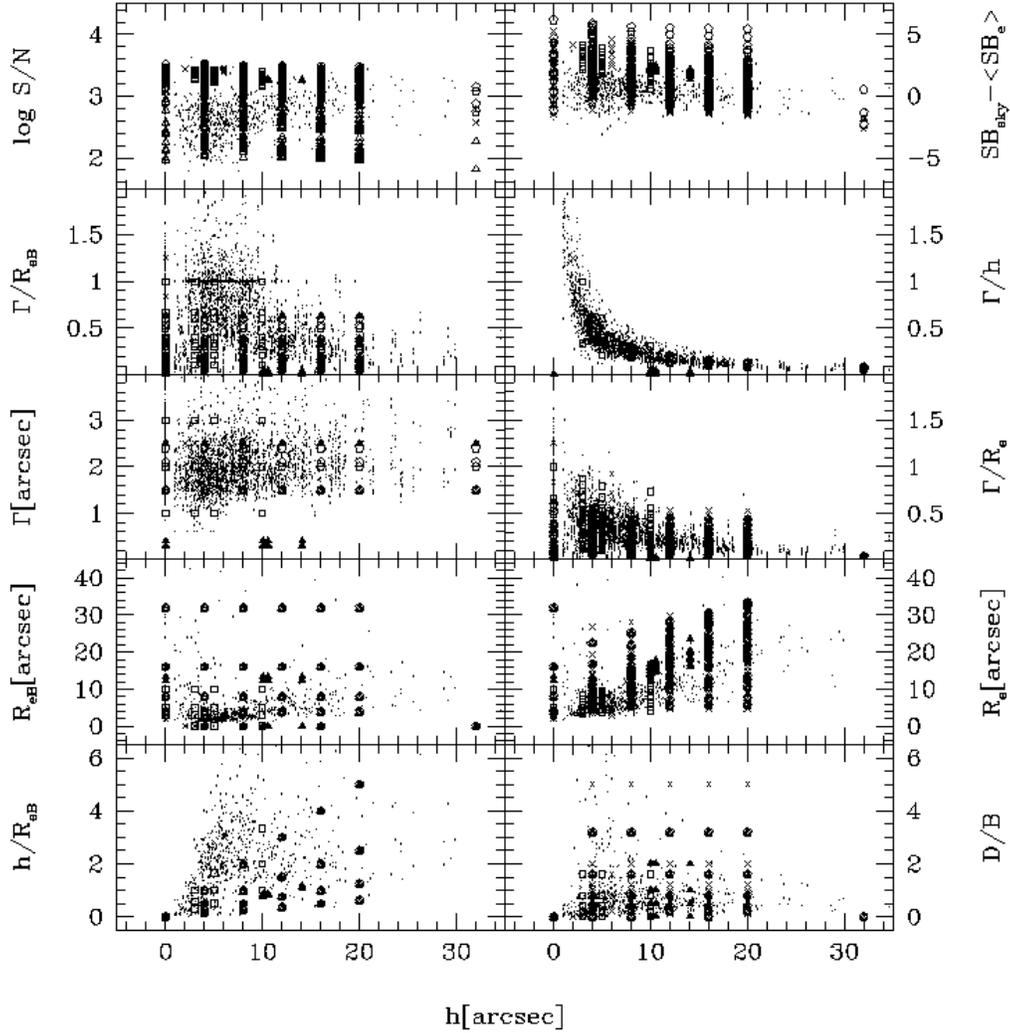}
\caption[f1.eps]{The parameter space of the $R^{1/4}$ plus exponential
profile of the Monte Carlo simulations discussed in Figure
\ref{figresdb}-\ref{figgamma}.  Models of
Figs. \ref{figresdb}-\ref{figskyfix}: crosses (see also Table 2). Models of
Fig. \ref{figextension}: skeletal triangles. Models of
Fig. \ref{figflux}: open triangles. Models of
Figs. \ref{figseeing}-\ref{figgamma}: open squares. Models of
Fig. \ref{figcombination}: open pentagons. Models of Fig.
\ref{figbulgedisk}: open hexagons.  The small dots show the
position of the EFAR galaxies as determined in Paper 3. The parameters
of bulge only models are shown with $h=0$. The parameters of disk only
models are shown at $R_{eB}=0$. See discussion in \S \ref{montecarlo}.}
\label{figparspace}
\end{figure}

In this section we discuss the results obtained by fitting the models
indicated by the crosses in Figure \ref{figparspace}. For clarity the
parameters are also given in Table \ref{tabpara}.
No sky subtraction
errors are introduced and the sky correction algorithm is not used. The
detailed analysis of the possible sources of systematic errors
discussed in \S \ref{sky}-\ref{snratio} is performed on the same 
sample of  models. More extreme values of the
parameters are used when testing the effects of seeing and resolution 
(\S \ref{seeing}).  The profiles tested in this section extend out to 
$4 R_e$, have a pixel size of 0.4 arcsec and normalization of $10^7$
counts, with $G_i=RON_i=1$ (see Eq. \ref{weightw}),
corresponding to $S/N\approx 1000$.

%\placetable{tabpara}
\begin{table}[hp]
\centering
\caption[*]{The parameters of the models indicated by the crosses of
Figure \ref{figparspace} (see \S \ref{parameter}). A model for each
combination of parameters in the two blocks separately has been generated. 
$D/B=\infty$ indicates exponential models ($B/D=0$). } 
\label{tabpara}
\begin{tabular}{ll}
&\\
\tableline
\tableline
Parameter & Values \\
\tableline
&\\
$R_{eB}('')$ & 4, 8, 12, 16, 20, 32\\
$h('')$       & 4, 8, 12, 16, 20, 32\\
$D/B$         & 0, 0.1, ..., 1, 1.2, 1.6, 2, 3.2, 5, $\infty$\\
$\Gamma('')$  & 1.5, 2.5\\
Sky/pixel     & 1000 \\
&\\
\tableline
&\\
$R_{eB}('')$ & 2, 3\\
$h('')$       & 3, 6\\
$D/B$         & 0, 0.2, 0.4, 0.8, 1.6, $\infty$\\
$\Gamma('')$  & 1.5, 2.5\\
Sky/pixel     & 500 \\
&\\
\tableline
\end{tabular}
\end{table}

\begin{figure}
\plotone{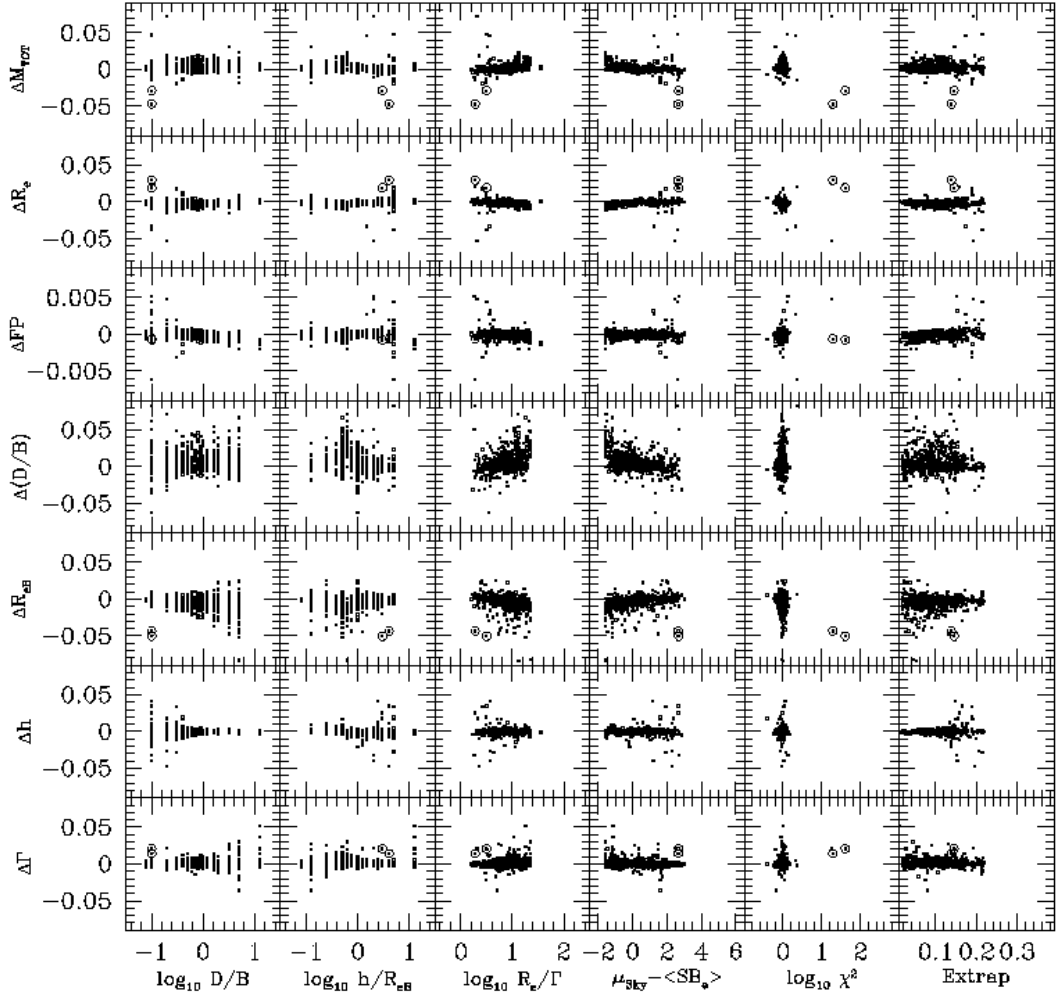}
\caption[f2.eps]{The reconstructed parameter space for the models 
indicated
by the crosses in Fig. \ref{figparspace}. No sky error is present.
The quantities plotted on the y-axis are defined as  $\Delta
M_{TOT}=M_{TOT}-M_{TOT}^f$, $\Delta R_e=\log R_e/R_e^f$, $\Delta FP =
\Delta R_e-0.3(\Delta M_{TOT}+5\Delta R_e)=\Delta R_e-0.3\Delta
\langle SB_e \rangle$, $\Delta (D/B)=\log [(D/B)/(D/B)^f$], $\Delta
R_{eB}=\log R_{eB}/R_{eB}^f$, $\Delta h = \log h/h^f$, $\Delta \Gamma
= \log \Gamma/\Gamma^f$. On the x-axis, the first three boxes show the
input parameters of the models in the logarithm units ($\log D/B$,
$\log h/R_{eB}$, $\log R_e/\Gamma$). The last three boxes show the
differences in magnitudes between the assumed sky value and the
average effective surface brightness of the models ($\mu_{\hbox{\rm
Sky}}-\langle SB_e \rangle$), the logarithm of the reduced $\chi^2$,
and the fraction of light extrapolated beyond $R_{max}$ used in the 
determination of
$M^f_{TOT}$. Models with $D/B=0$ (pure $R^{1/4}$ laws) are plotted at
$\log D/B=-1.1$ and $\log h/R_{eB}=-1.1$. Models with $B/D=0$ (pure
exponential laws) are plotted at $\log D/B=1.1$ and $\log
h/R_{eB}=1.1$. Models with $D/B\neq 0$ which have been fitted with one
component are circled.  See \S \ref{parameter} for a discussion of
the results.}
 \label{figresdb}
\end{figure}

\begin{figure}
\plotone{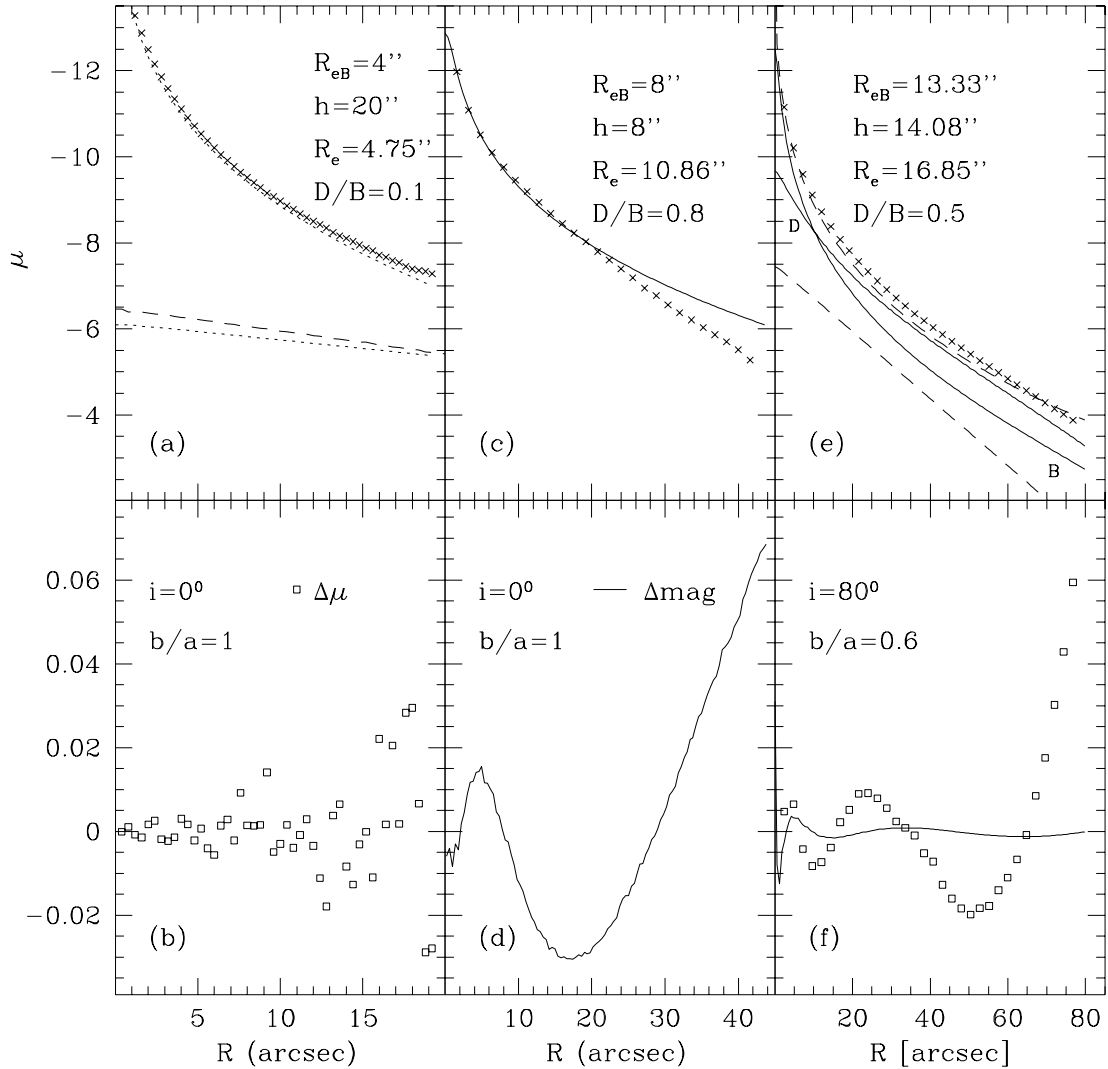}
\caption[f3.eps]{(a)  A circular disk plus bulge model with $D/B=0.1$ 
and $h/R_{eB}=5$ 
(crosses). The dotted curves show the luminosity profiles
$\mu(R)=-2.5\log I(R)$ of the bulge and the disk components, the dashed
curve the fitted disk component. (b) The differences $\Delta \mu$ (in
mag arcsec$^{-2}$, open squares) between  
model surface brightness and the fitted one (dotted curve) (see \S 
\ref{parameter}). 
(c)  The $R^{1/4}$  fit (solid curve) to the surface brightness 
magnitude profile of a circular disk plus bulge model with $D/B=0.8$ 
and $h/R_{eB}=1$ (crosses, one point in every
four). (d) The differences $\Delta mag$ between the $R^{1/4}$ 
integrated 
magnitudes and the fitted ones (solid curve, see \S \ref{parameter}). Note 
that $|\Delta mag| <0.07$ even if large deviations $\Delta \mu$ are present.
(e) The fit to the circularized profile of a flattened bulge plus an
inclined disk
model (see \S. \ref{decomposition}). The luminosity profile of the model 
(crosses, one point in every four; the bulge
and the disk components, with the listed parameters, are the full
curves) is best fitted by an $R^{1/4}$ plus exponential law (dashed
curves) with parameters $R_{eB}^f=16.34$ arcsec, $h^f=13.93$ arcsec,
 $(D/B)^f=0.13$, $R_e=17.51$ arcsec.
(f) The residuals $\Delta \mu$ of the fit (open squares, one point in every
four) and the differences 
between the growth curves $\Delta mag$ (full curve).}
 \label{figbadfit}
\end{figure}

Figure \ref{figresdb} shows the precision of the reconstructed
parameters. Total
magnitudes are derived with a typical accuracy of 0.01 mag, $R_e$ and
$\Gamma$ to 3\%, $R_{eB}$ and $h$ to $\approx 8$\%, $D/B$ to $\approx
10$\%. The errors $\Delta M_{TOT}=M_{TOT}-M_{TOT}^f$ and $\Delta R_e
=\log R_e/R_e^f$ are highly
correlated, with insignificant differences from the relation $\Delta FP =
\Delta R_e-0.3(\Delta M_{TOT}+5\Delta R_e)= \Delta R_e-0.3\Delta
\langle SB_e \rangle$. Galaxies with faint ($D/B<0.3$) and shallow
($h/R_{eB}>2$) disks show the largest deviations. This partly reflects
a residual (minimal) inability of the fitting program to converge to
the real minimum $\chi^2$ (there are 3 points with $\chi^2>10$), but stems
also from the degeneracy of the Bulge plus Disk fitting. Figure
\ref{figbadfit} (a) and (b) show the example of a model with $D/B=0.1$
and $h/R_{eB}=5$ where a very good fit is obtained ($\chi^2=1.3$) yet
there is a 0.05 mag error on $M_{TOT}$ and the disk solution is
significantly different from the input model.  Note that the largest deviations
$\Delta M_{TOT}$ and $\Delta R_e$ are associated with the largest
extrapolations ($\approx 20$\%). In the following sections we shall
see that extrapolation is the main source of uncertainty, when
determining total magnitudes and half-luminosity radii. The
uncertainties $\Delta R_{eB}$ on the bulge scale length are smallest
with bright bulges, while those on the disk scale length $\Delta h$ are
smallest with bright disks. The algorithm to opt for one-component
best-fits (Eqs. \ref{chibo}-\ref{chido}) identifies successfully all of
the one-component models tested (bulges plotted at $\log D/B =-1.1$
and $\log h/R_{eB}=-1.1$, disks plotted at $\log D/B =1.1$ and $\log
h/R_{eB}=1.1$ in Figure \ref{figresdb}). For only two models (with
$D/B=0.1$ and large $h/R_{eB}$) is the bulge-only fit preferred (using
the 3$\sigma$ test) to the
two-component fit (circled points in Figure \ref{figresdb}).

\begin{figure}
\plotone{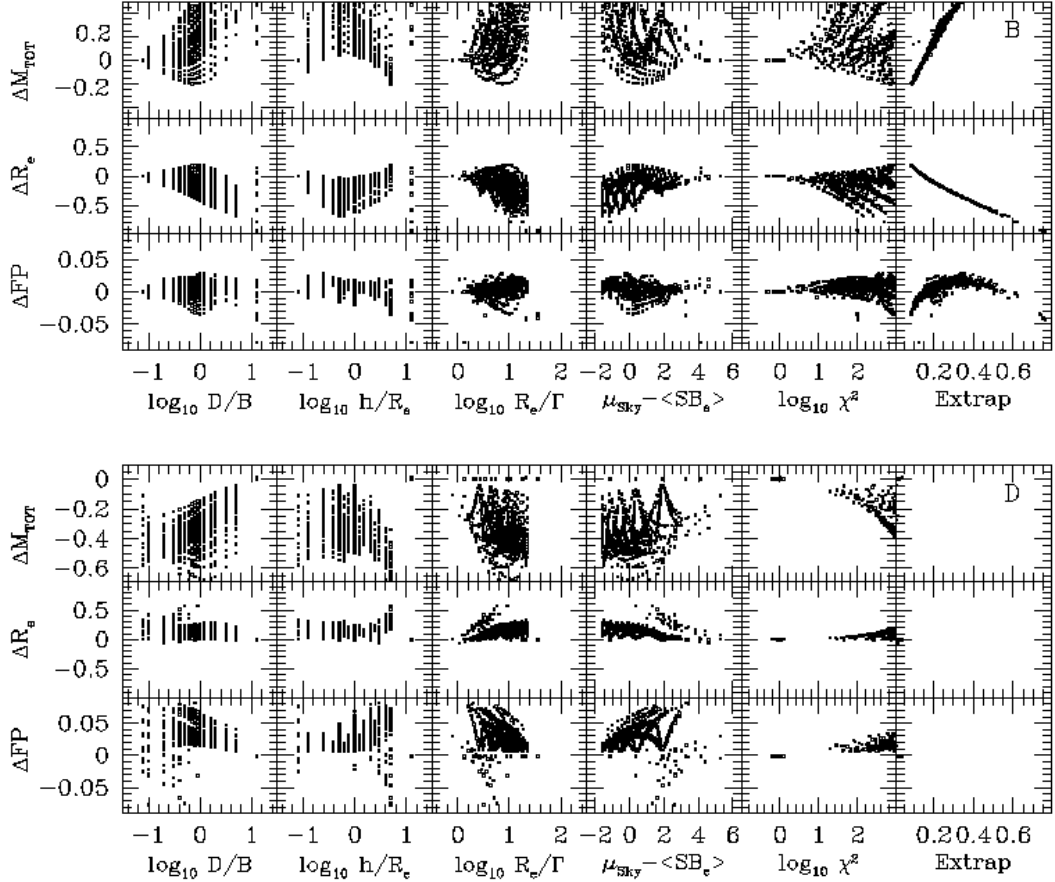}
\caption[f4.eps]{The effects of fitting disk plus bulge test
 profiles by either 
a single bulge (first three rows of plots) or a single disk (last three rows 
of plots) model.
The test models are indicated by the crosses of
Fig. \ref{figparspace}. $\Delta M_{TOT}$, $\Delta R_e$, and $\Delta FP$ 
are defined as in Fig. \ref{figresdb}, x-axis as in Fig. 
\ref{figresdb}. See \S \ref{parameter} for the a discussion of the results.
Note the change of scale on the ordinate axis with respect to Figure 
\ref{figresdb}.}
\label{figquatexp}
\end{figure}

Figure \ref{figquatexp} shows the results obtained by fitting a pure
bulge or a pure disk. As before, no sky subtraction error is introduced 
and the sky correction algorithm is not activated. Neglecting one of the
 two components strongly
biases the derived total magnitudes and half-luminosity radii. In the
case of the $R^{1/4}$ fits, already test models with values of D/B as small as
$\approx 0.2$ give fitted 
magnitudes wrong by 0.2 mag, and $R_e$ by more than 30\%. The
systematic differences correlate with the amount of extrapolation
involved, and large extrapolations yield strongly overestimated
magnitudes and half-luminosity radii.  However, the resulting
correlated errors $|\Delta FP|$ are almost always smaller than 0.03.  In the
case of pure exponential fits, the derived total magnitudes and
half-luminosity radii are always smaller than the true values, since
very little extrapolation ($<1$\%) is involved. Consequently, a
positive, correlated error $\Delta FP$ ($\approx +0.03$) is
obtained. Finally, note that pure bulge fits are {\it bad} fits of the
surface brightness profiles ($\chi^2>10$), but may appear to give
acceptable fits of the integrated magnitude profiles. (One can easily
show that the differences in integrated magnitudes are the weighted
mean of the differences in surface brightness magnitudes).  Figure
\ref{figbadfit} (c) and (d) shows such an example for an $R^{1/4}$ fit
to a model with $D/B=0.8$ and $h/R_{eB}=1$. The residuals in the
integrated magnitude profile are always smaller than 0.07 mag, but a
$\chi^2=181$ is derived, with $\Delta M_{TOT}=0.32$ and
$R_e^f/R_e=1.65$.  These considerations suggest that magnitudes and
half-luminosity radii derived by fitting the $R^{1/4}$ curve of growth
to integrated magnitude profiles (Burstein et al. 1987, Lucey et
al. 1991, J\o rgensen et al. 1995, Graham 1996) may be subject to systematic
biases, as indeed Burstein et al. (1987) warn in their Appendix. 
This might be important for the sample of 
J\o rgensen et al. (1995), where substantial disks are detected in a large
fraction of the galaxies by means of an isophote shape analysis. It is 
certainly very important for
the sample of cD galaxies studied by Graham (1996, see discussion in
Paper III).
These objects have luminosity profiles which differ strongly from an $R^{1/4}$
law. 

\begin{figure}
\plotone{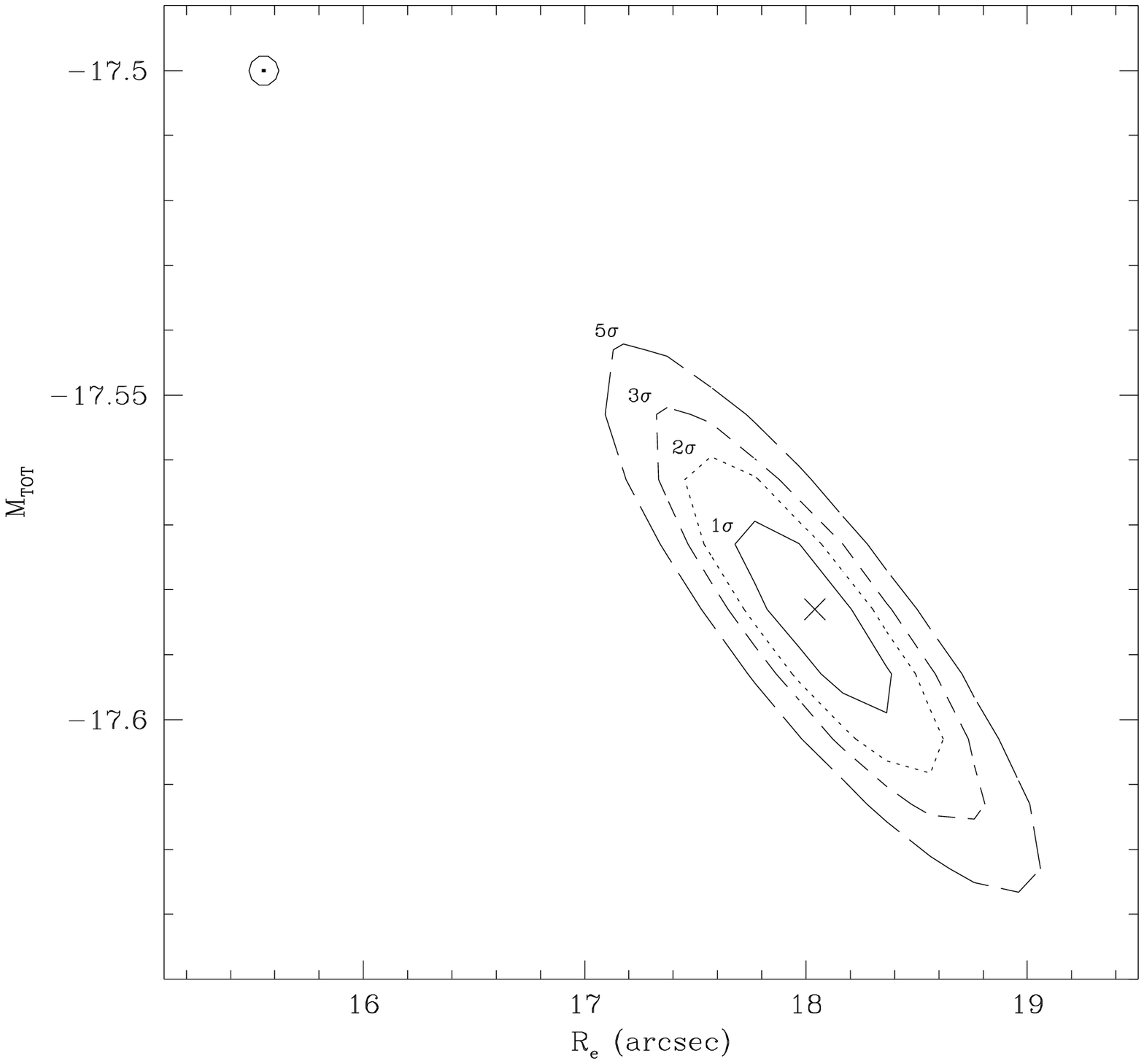}
\caption[f5.eps]{Illustration of the underestimation of the errors. 
The contours of constant $\chi^2$ near the minimum
of an $R^{1/4}$ fit to a $h/R_{eB}=0.5$, $D/B=0.1$ disk plus bulge model. 
The cross shows the best-fit solution, the circle near the upper left corner 
gives the real parameters
of the model. The errors estimated at
the $5\sigma$ contour underestimate the differences between the model
and the fit by a factor 2 (see 
\S \ref{parameter}).}
\label{figcontour}
\end{figure}

Finally, note that the {\it systematic} errors shown in Figure
\ref{figquatexp} (and in the figures of the following sections) cannot
be simply estimated by considering the shape of the $\chi^2$ function
near the minimum.  Figure \ref{figcontour} shows the $1,2,3,5\sigma$
contours of constant $\chi^2$ for an $R^{1/4}$ fit to a
$h/R_{eB}=0.5$, $D/B=0.1$ $R^{1/4}$ plus exponential model. The
reduced $\chi^2$
(8.47 at the minimum) has been normalized to 1, so that $1\sigma$
corresponds to a (normalized) $\chi^2=1+\sqrt{2/N_{free}}=1.11$.  The
errors, estimated at the $5\sigma$ contour, underestimate the
differences between the fit and the model by a factor 2.  This results from
the extrapolation involved and can be as large as one order of magnitude
for models with larger D/B ratios.

%\placefigure{figparspace}
%\placefigure{figresdb}
%\placefigure{figbadfit}
%\placefigure{figquatexp}
%\placefigure{figcontour}
%\newpage
%

\subsection{Sky subtraction errors}
\label{sky}
\begin{figure}
\plotone{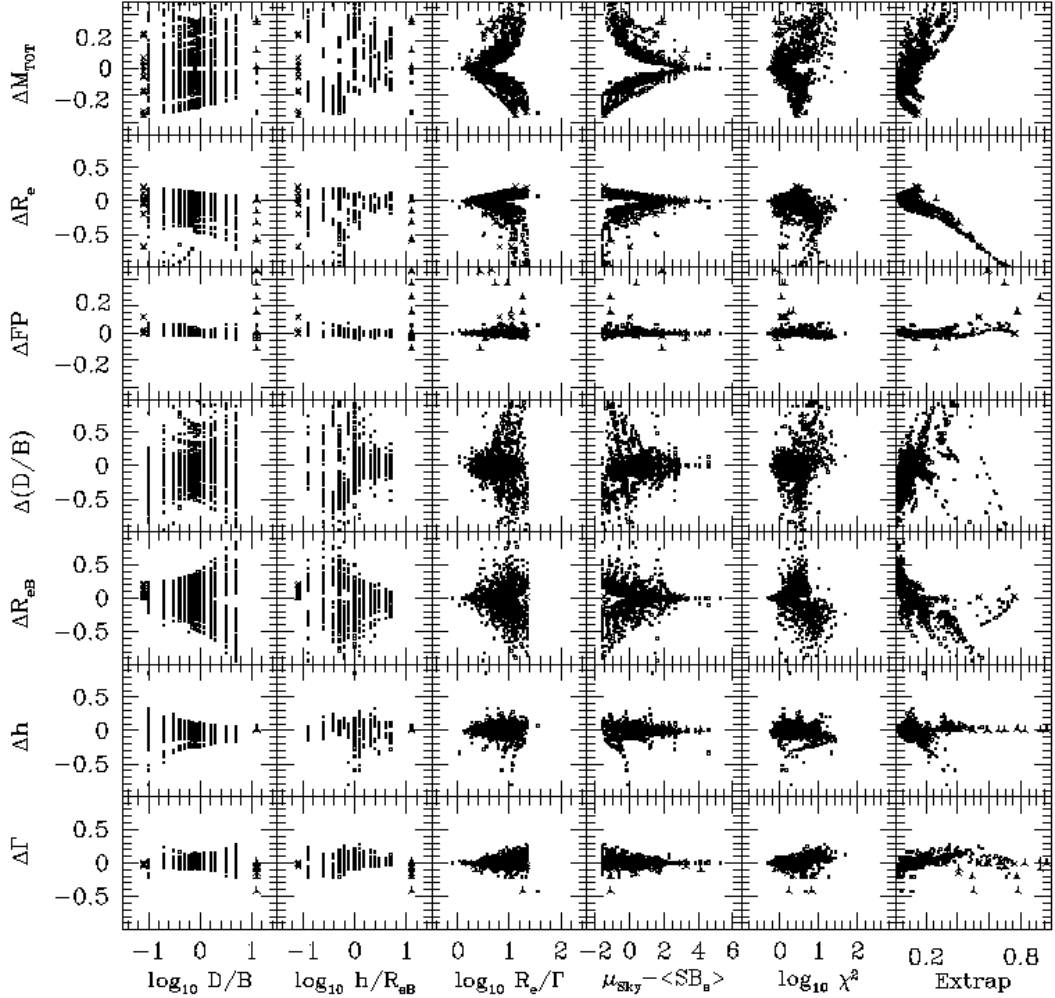}
\caption[f6.eps]{The biases introduced by a $\pm 1$\% sky subtraction 
error. 
Quantities plotted as in Fig. \ref{figresdb}.  Models with $D/B=0$ which 
have been fitted with two components are shown as crosses.
Models with $B/D=0$ which have 
been fitted with two components are shown as triangular crosses.
Note the change of scale on the ordinate axis with respect to 
Figure \ref{figresdb}. See discussion in \S \ref{sky}.}
\label{figskyerror}
\end{figure}

Sky subtraction errors can induce severe systematic errors on the
derived photometric parameters of galaxies.  Figure \ref{figskyerror}
shows the parameters derived from the $R^{1/4}$ plus exponential
models examined in the previous section, where now the sky has been 
overestimated or underestimated by  $\pm 1$\%. The sky correction 
algorithm is not activated. 

The biases become increasingly large as the sky brightness approaches
the effective surface brightness of the models. As expected, 
underestimating  the sky (a negative sky error) produces 
total magnitudes that are too bright  and 
half-luminosity radii that are too large relative to  the true ones. 
 The size of
the bias correlates with the extrapolation needed to derive
$M_{TOT}^f$. The opposite happens when the sky is overestimated, but
the amplitude of the bias is smaller, because there is no extrapolation. 
The correlated error $\Delta FP$ remains small ($\approx
0.05$), except for the cases where large extrapolations are
involved. The D/B ratio is ill determined, with better precision for
models with extended disks ($h/R_{eB}>2.5$). The scale length of the
bulge is better determined for low values of $D/B$ (dominant bulge),
the scale length of the disk component is better determined for large
values of $D/B$ (dominant disk). The parameter least affected  is the
value $\Gamma$ of the seeing, which is determined in the inner, bright
parts of the models, where sky subtraction errors are unimportant. 
Bulge-only or disk-only models appear to be fit best 
by two-component models (crosses and triangular crosses in Figure
 \ref{figskyerror}).
Finally, note that reasonably good fits ($\chi^2<10$)
to the surface brightness profiles are always obtained, in spite of
the large errors on the reconstructed parameters.

\begin{figure}
\plotone{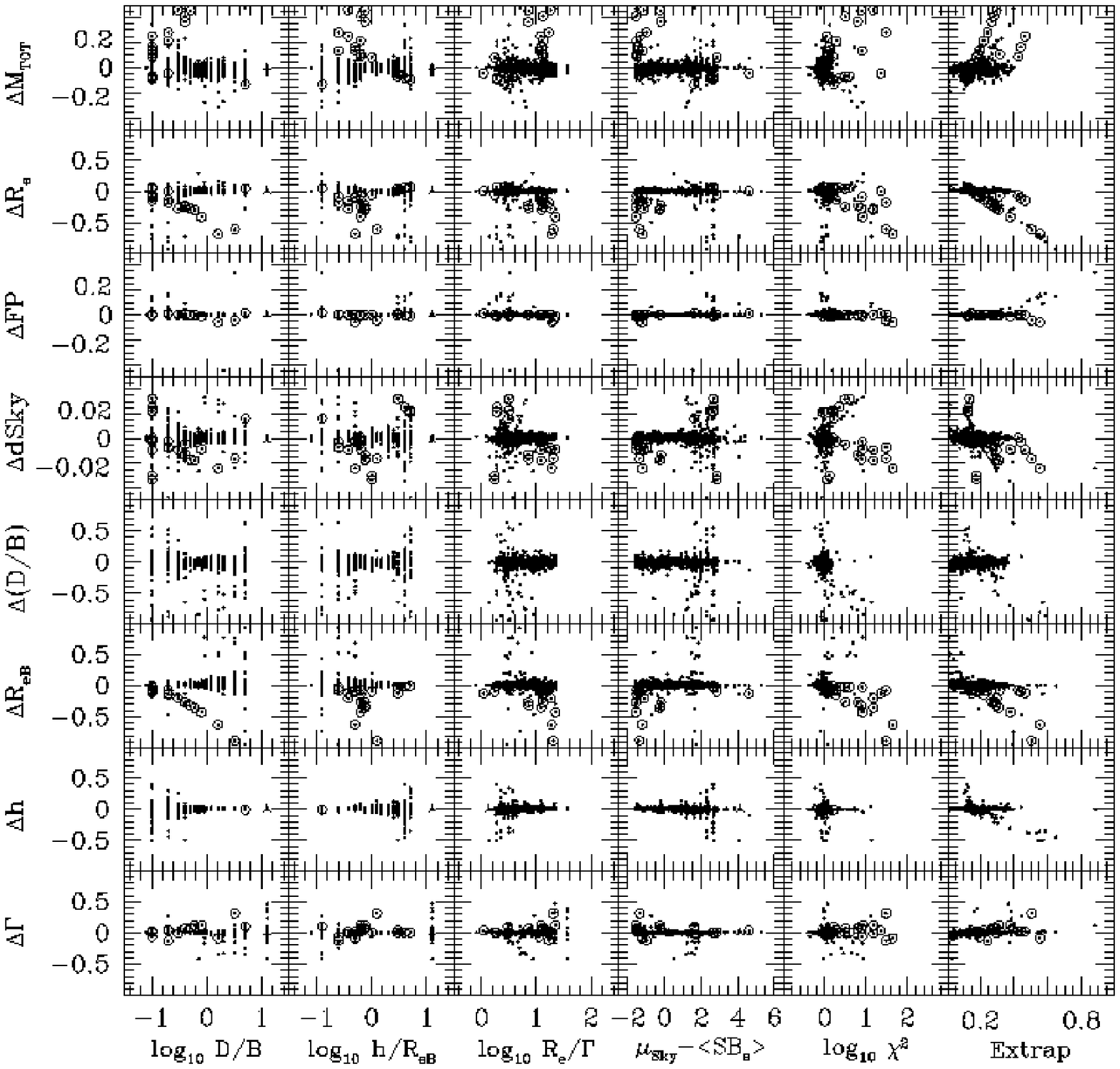}
\caption[f7.eps]{The effects of the sky fitting algorithm. 
The parameters of the models of 
Fig. \ref{figskyerror} with the sky subtraction errors of 0, $\pm1$\%, 
$\pm3$\%, are reconstructed using the sky fitting algorithm. Quantities
and symbols plotted as in Figures \ref{figresdb} and \ref{figskyerror}.
In addition, the difference
$\Delta$dSky=dSky/Sky-dSky$^f/$Sky on the sky correction is plotted.
Note the change of scale on the ordinate axis with respect to Figure 
\ref{figresdb}. See discussion in \S \ref{sky}.}
 \label{figskyfix}
\end{figure}

The biases discussed above can be fully corrected when the sky-fitting 
algorithm of Eq. \ref{chidmax} is applied. Figure \ref{figskyfix} shows 
the reconstructed parameters of the models considered in \S \ref{parameter},
where sky subtraction errors of 0, $\pm1$\%, $\pm3$\% have been introduced.
For most of the models examined, the errors on the derived quantities are no
more than a factor 2 larger than those shown in Fig. \ref{figresdb}. The sky 
corrections are computed to better than 0.5\% precision. 
Larger errors $\Delta M_{TOT}$ and $\Delta R_e$ are obtained for models with
relatively weak  ($D/B<0.3$) and extended disks ($h/R_{eB}>2.5$), where the 
degeneracy
discussed in \S \ref{parameter} is complicated by the sky subtraction 
correction. These cases give reasonably good fits ($\chi^2<10$), but are identified by
the large extrapolation ($>0.3$) involved. Models with concentrated disks 
($h/R_{eB}<0.2$) can also be difficult to reconstruct, when $h/\Gamma\approx 
1$. For some of these problematic fits, one-component solutions are preferred
by Eqs. \ref{chibo}-\ref{chido} (circles in Eq. \ref{figskyfix}).

%\placefigure{figskyerror}
%\placefigure{figskyfix}
%

A common problem of CCD galaxy photometry is
the relatively small field of view, particularly with the older smaller CCDs. 
If the size (projected on the sky) 
of the CCD is not large enough compared to the half-luminosity radius of
the imaged galaxy, then the sky as determined on the same frame
will be contaminated by galaxy light and biased to values larger 
than the true one. Total
magnitudes and half-luminosity radii can therefore be biased to smaller
values, the effect being more important for intrinsically large galaxies,
which tend to have low effective surface brightnesses. 
The mean surface brightness in the annulus with radii
$R_i^{max}$ and $2R_i^{max}$ (see \S \ref{diskbulge}) predicted by the fit 
allows us to estimate the size of the contamination. 

\subsection{Radial extent}
\label{extension}

\begin{figure}
\plotone{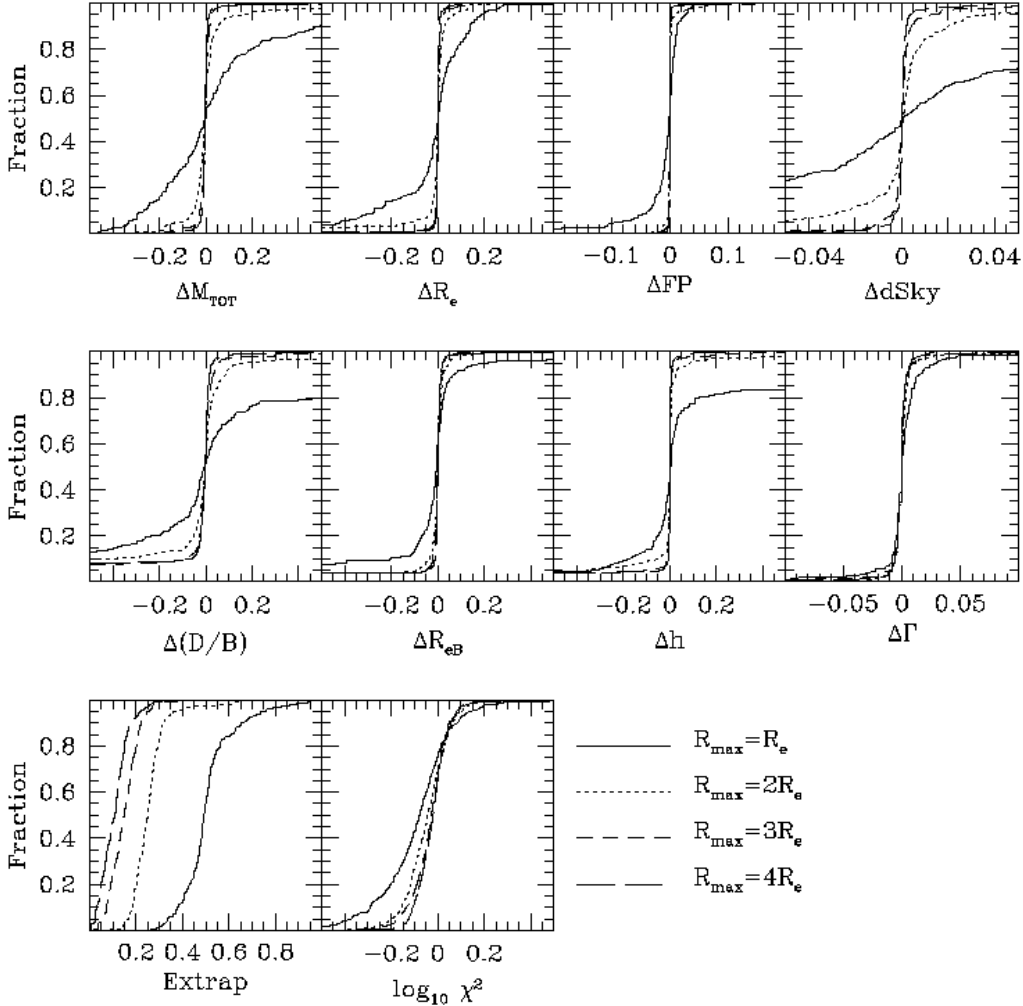}
\caption[f8.eps]{The effect of the radial extent of the profiles 
on the precision
of the derived parameters. The cumulative distributions of the errors on the 
derived photometric
parameters as derived from the simulations are shown for a range of $R_{max}$
 values (full lines: $R_{max}=R_e$, dotted lines: $R_{max}=2R_e$, dashed lines
$R_{max}=3R_e$, long-dashed lines: $R_{max}=4R_e$). Good reconstructions are 
obtained when $R_{max}/R_e>2$ (see \S \ref{extension}).}
\label{figextension}
\end{figure}

Photoelectric photometry of large, nearby galaxies rarely goes beyond
1 or 2 $R_e$ (Burstein et al. 1987) and the same applies for the surface
photometry obtained with smallish CCDs.  The typical profiles obtained
in Paper III extend to a least 4 $R_e$, but a small fraction of them are
less deep, reaching 1 or 2 $R_e$ only. Here we investigate the effect of
the radial extent of the profiles, keeping the normalization of the
profiles fixed ($10^7$ counts, $S/N\approx 10^3$).  
Sky subtraction errors of $0$,
$\pm3$\% are introduced and the sky fitting is activated. Figure
\ref{figextension} shows the cumulative distributions of the errors on the 
derived photometric
parameters as derived from the simulations, for a range of $R_{max}$ values. 
When $R_{max}=R_e$, rather
large errors are possible (0.3 mag in the total
magnitude, $>30$\% in $R_e^f$). 
The main source of error is again the large extrapolation
involved when $R_{max}\approx R_e$, coupled with the sky correction
which becomes unreliable for these short radial extents.  As soon
as $R_{max}\ge3R_e$ the errors reduce to the ones discussed in
\S \ref{parameter}. The same kind of trend is observed for the
parameters of the two component ($\Delta (D/B)$, $\Delta R_{eB}$,
$\Delta h$). The seeing values are less affected, as they are
sensitive to the central parts of the profiles only. Finally, note that
in all cases very good fits are obtained ($\chi^2\approx 1$).

%\placefigure{figextension}
%

\subsection{Signal-to-noise ratio}
\label{snratio}

\begin{figure}
\plotone{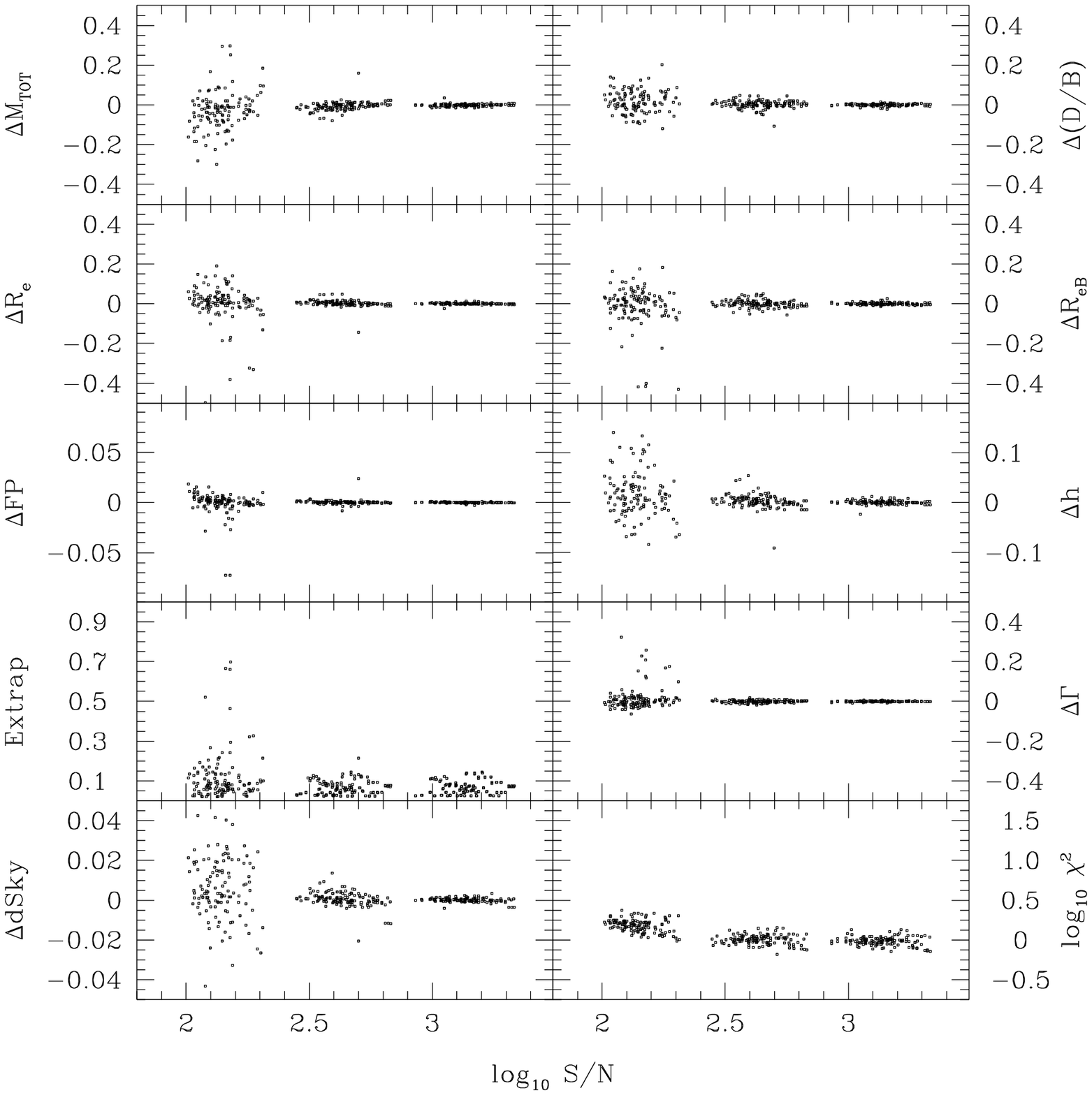}
\caption[f9.eps]{The effect of the signal-to-noise ratio of the 
profiles on the
precision of the derived parameters. Good reconstructions are obtained
when $S/N>300$ (see \S \ref{snratio}).
Note the change of scale on the ordinate axis with respect to Figure 
\ref{figresdb}.}
\label{figflux}
\end{figure}

For most of the galaxies discussed in Paper III, multiple profiles are
available with integrated signal-to-noise ratios $S/N> 300$, the
normalization used in the previous sections. But for some of the luminosity
profiles a smaller number of total counts has been collected (see Figure
\ref{figparspace}).  Here we
investigate how the signal-to-noise ratio of the profiles affects the
outcome of the fits.  As
before, the subset of models of \S \ref{sky} is used with
$R_{max}\le4 R_e$ (see comment at the beginning of
\S \ref{montecarlo}).  Sky subtraction errors of $0$, $\pm3$\% are
introduced and the sky fitting is activated.  Figure \ref{figflux}
shows how the errors on the derived parameters increase when the
signal-to-noise ratio is reduced.  For fluxes as low as about $10^5$
($S/N\approx 10^2$) all of the derived photometric parameters become
uncertain (0.2 mag in the total
magnitudes, 20\% variations in the derived $R_e$, large spread $\Delta (D/B)$,
 $\Delta R_{eB}$, $\Delta h$,
$\Delta \Gamma$), as large extrapolations and uncertain sky
corrections are applied. In all cases very good fits are obtained 
($\chi^2\approx 1$).

%\placefigure{figflux}
%

\subsection{Seeing and sampling effects}
\label{seeing}

\begin{figure}
\plotone{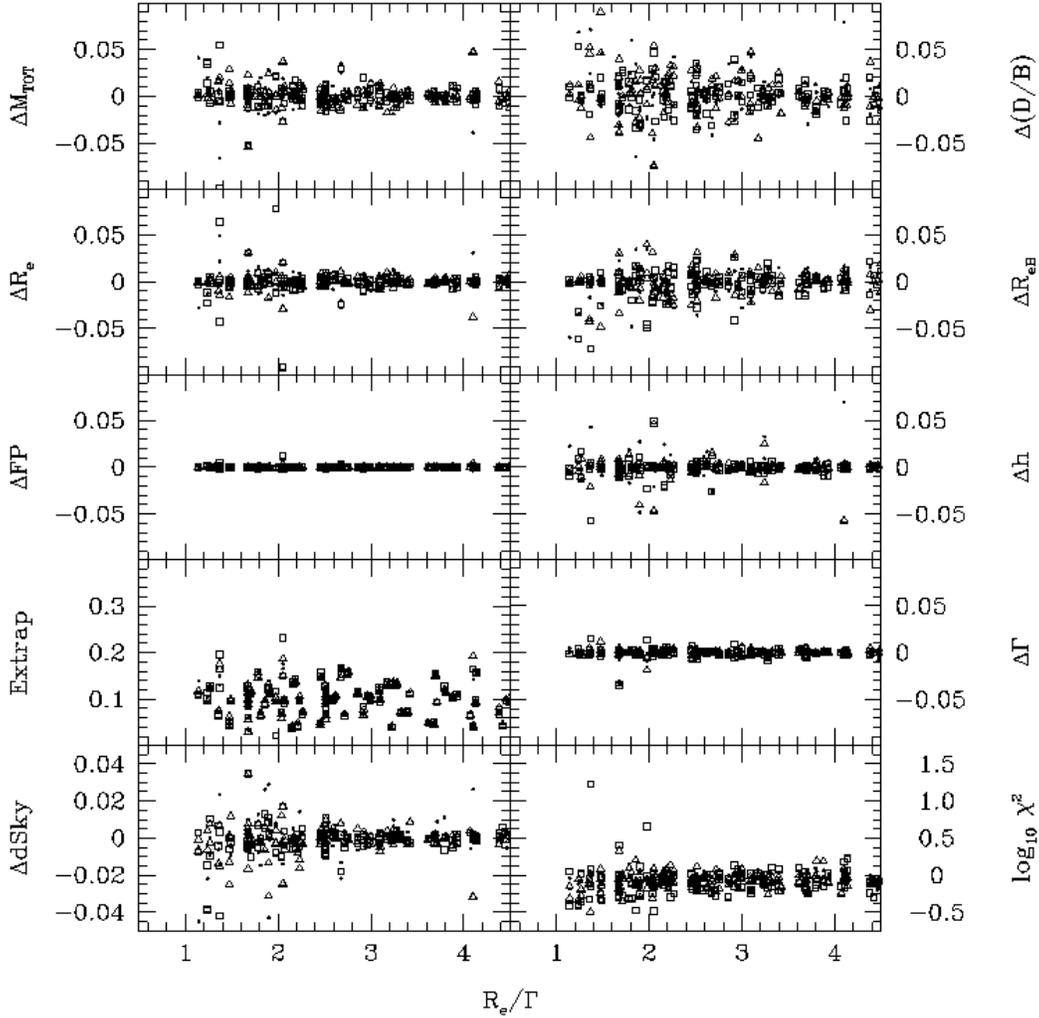}
\caption[f10.eps]{The effect of seeing and pixel sampling of the 
profiles on the 
precision of the derived parameters. Different symbols indicate different
pixel sizes (small dot 0.4 arcsec, triangles 0.6 arcsec, squares 0.8 arcsec).
Note the expanded ordinate scale with respect to Figures 
\ref{figquatexp}-\ref{figflux}. See discussion in \S \ref{seeing}.}
\label{figseeing}
\end{figure}

Some of the galaxies considered in Paper III are rather small, with
$R_e<4''$.  Here we investigate the effects of seeing and pixel
sampling, when $R_e\approx\Gamma\approx$pixel size. Figure
\ref{figseeing} shows that reliable parameters can be derived down to
$R_e\approx\Gamma$, with  pixel sizes  0.4-0.8 arcsec, with only a
small increase of the scatter for $R_e<2\Gamma$. 

\begin{figure}
\plotone{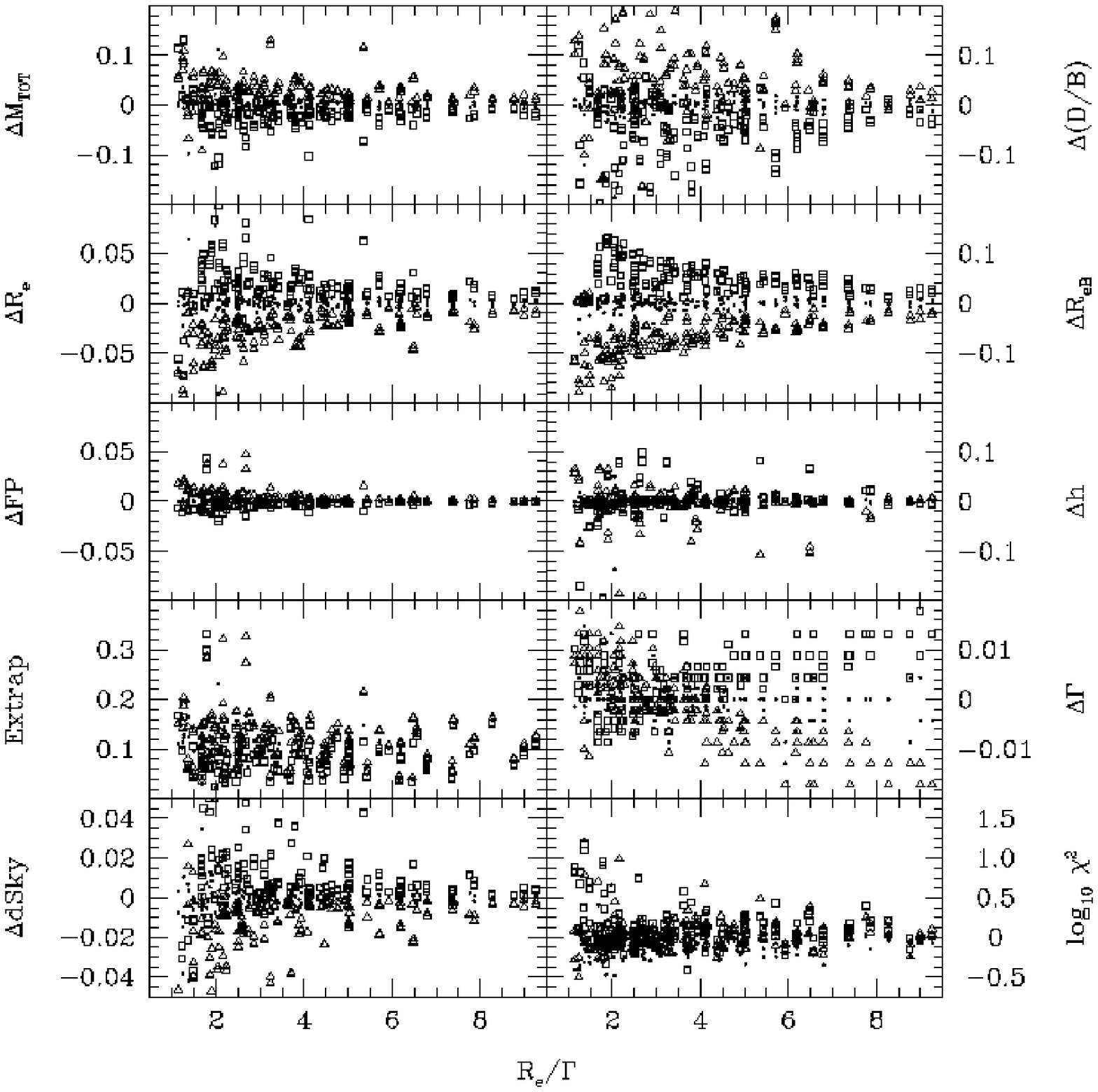}
\caption[f11.eps]{The effect of the choice of the psf on the 
precision of the derived parameters. Open triangles for $\gamma=1.5$, dots for
$\gamma=1.6$ and open squares for $\gamma=1.7$. Fits performed with the 
$\gamma=1.6$ psf overestimate (underestimate) magnitudes and half-luminosity
radii of models constructed with $\gamma=1.5$ ($\gamma=1.7$; see \S 
\ref{seeing}). Note the expanded ordinate scale with respect to Figures 
\ref{figquatexp}-\ref{figflux}.}
\label{figgamma}
\end{figure}

A small systematic effect is caused by the choice of the psf. Saglia et al.
(1993) demonstrate that a good approximation of the psfs observed during
the  runs described in Paper III is given by the $\gamma$ psf
with $\gamma=1.5-1.7$. We adopt $\gamma=1.6$ for the fits. Here 
we test the effect of having $\gamma=1.5$ or 1.7 with a pixel size of 
0.8 arcsec. Figure \ref{figgamma} shows that if $\gamma=1.5$ is the true
psf of the observations, then the half-luminosity radius,
the total luminosity, the scale length of the bulge will be slightly 
overestimated, and the disk to bulge ratio slightly underestimated.  
A small systematic trend is observed in the correlated errors $\Delta FP$.
The scale length of the disk component is less affected. The sky 
corrections are also biased, but do not strongly affect  the photometric 
parameters, because of the high average surface brightness of the small 
$R_e$ models. Seeing values suffer a very small, but systematic effect.
The opposite trends are observed if the true $\gamma$ is 1.7. 
In all cases very good fits are obtained 
($\chi^2\approx 1$). The systematic differences
become unimportant for $R_e>2\Gamma$. 

Finally, the  seeing values derived can be systematically biased, if
the central concentration of the fitted galaxies does not match the
one of the $R^{1/4}$ plus exponential models. We investigate this
effect by fitting the $\Psi=12$ plus exponential or the smoothed
$R^{1/4}$ plus exponential models discussed in
\S \ref{diskbulge}. We find that in the first case the seeing
value is underestimated which compensates for the higher concentration of
the $\Psi=12$ component. The shallow radial decline of the luminosity
profile in the outer parts introduces systematic biases in the
reconstructed parameters, similar to those discussed for the $R^{1/n}$
profiles, for large values of $n$ (see \S \ref{profiles}).  The
half-luminosity radii and total magnitudes derived are underestimated 
 by 20\% and 0.2 mag respectively, when a $\Psi=12$ model
with no exponential component is fitted. The biases are reduced when
models with an exponential component are constructed.  In the case of
the smoothed $R^{1/4}$ law, the seeing value is overestimated to fit
the lower concentration
of the smoothed $R^{1/4}$ component. No biases are introduced on the
other reconstructed parameters.

%\placefigure{figseeing}
%\placefigure{figgamma}
%

\subsection{Tests of profile combination}
\label{testcombination}

\begin{figure}
\plotone{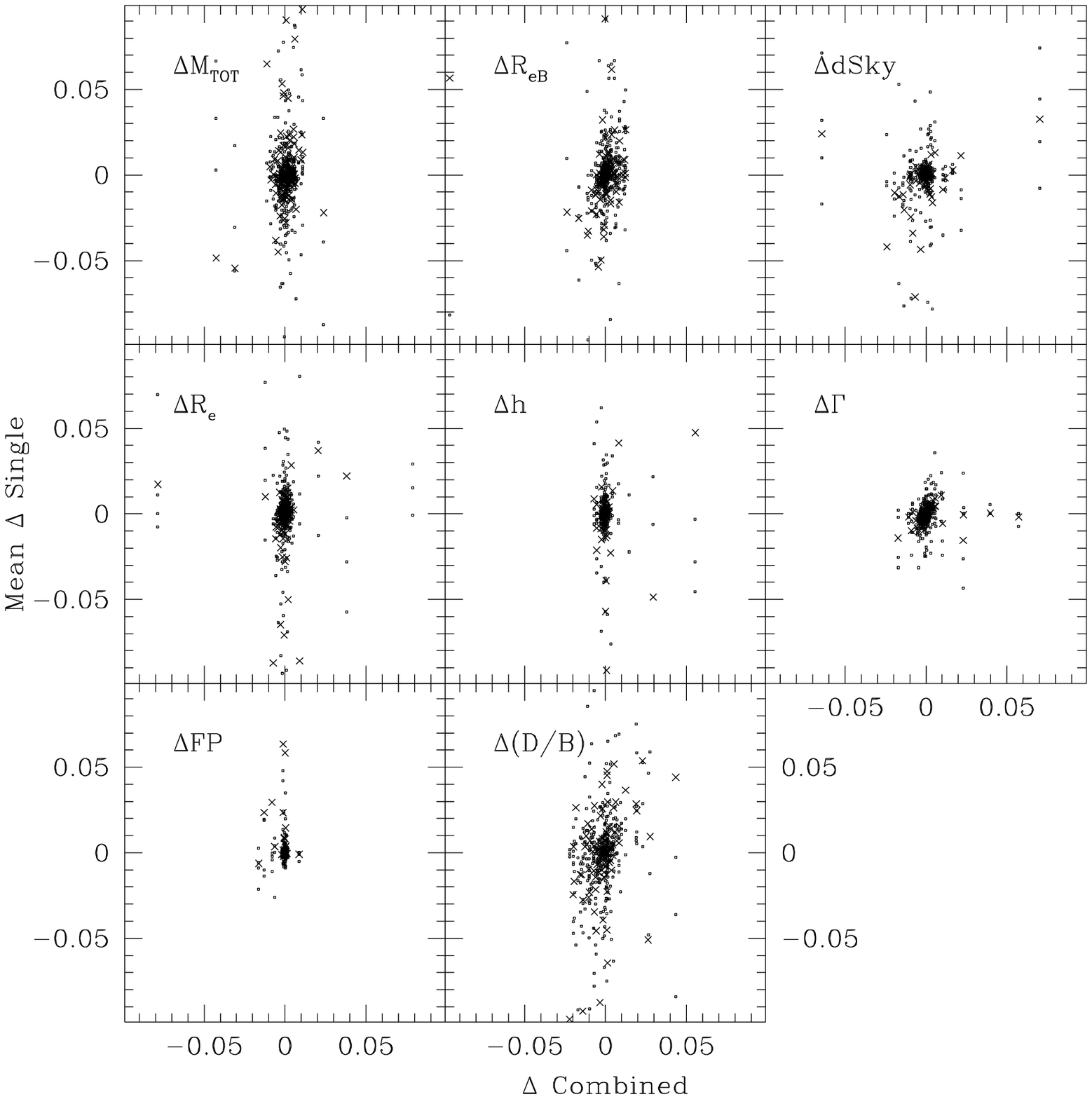}
\caption[f12.eps]{The profile combination algorithm and the 
precision of the derived parameters. The x-axis plots the
residuals $\Delta$ of the parameters derived using the fitting
procedure with profile combination. $\Delta$ dSky and $\Delta
\Gamma$ are averaged over the four obtained values.  The y-axis plots
the {\it mean} of the residuals of the parameters derived by fitting
each single independently as crosses, and the residuals of each fit as dots 
(see discussion in \S \ref{testcombination}).}
\label{figcombination}
\end{figure}

In order to test the combination algorithm described in
\ref{combination}, four profiles with different $\Gamma$, pixel sizes,
normalizations, gain, readout noise, and sky subtraction errors (see
Table \ref{tabcombination}; these parameters match the typical
values of the profiles of Paper III) are generated for the set of
models identified
by the open pentagons of Figure \ref{figparspace}.  Figure
\ref{figcombination} shows the result of the test. The abscissa plots the
residuals $\Delta$ of the parameters derived using the fitting
procedure with profile combination. $\Delta$ dSky and $\Delta
\Gamma$ are averaged over the four obtained values.  The ordinate plots
the {\it mean} of the residuals of the parameters derived by fitting
each single independently as crosses, and the residuals of each fit
as dots. The profile combination algorithm 
obtains  better precision on all of the parameters with the exception of 
$\Gamma$, where the maximum deviation is in any case smaller than 8\%.

%\placefigure{figcombination}
%

\clearpage
%\placetable{tabcombination}
\begin{table}[ph]
\centering
\caption[*]{The parameters of the multiple profiles test (see \S 
\ref{testcombination}).} 
\label{tabcombination}
\begin{tabular}{lrrrrr}
&&&&\\
\tableline
\tableline
Profile & 1 & 2 & 3 & 4 \\
\tableline
&&&&\\
Pixel size $('')$ & 0.4 & 0.606 & 0.862 & 0.792 \\
Sky per pixel  & 300 & 350 & 250 & 1500 \\
$\delta$Sky/Sky & $+1$\% & $-0.5$\% & $+1.5$\% & $+0.5$\% \\
$R_{max}/R_e$ & 4 & 3 & 4.5 & 2.5 \\
Normalization & $10^7$ & $5\times 10^6$ & $10^7$ & $5\times 10^6$ \\
Gain & 1 & 3 & 1 & 2 \\
Ron  & 1 & 4 & 1 & 5 \\
$\Gamma ('')$ & 2 & 2.1 & 1.5 & 2.4 \\
\tableline
\end{tabular}
\end{table}

\subsection{``Bulge'' and ``Disk'' components}
\label{decomposition}

The discussion of the previous sections shows that for a large fraction of the
parameter space, i.e. when deep enough profiles are available, with large
enough objects, not only can
the global photometric parameters $R_e$ and $M_{TOT}$ be reconstructed
with high accuracy, but also the parameters of the $R^{1/4}$ and the 
exponential components. Here we investigate further if reliable
``bulge'' and ``disk'' parameters can be derived, when the profiles
analysed are constructed  from the superposition of these two components.

With this purpose, we constructed a number of two-dimensional frames
(filled triangles in Figure \ref{figparspace})
as the sum of a flattened $R^{1/4}$ bulge and an exponential disk of
given inclination. The bulge (disk) frames follow
an exact $R^{1/4}$ (exponential) law with $R_{eB}=12\sqrt{b/a}$ arcsec
($h=10\sqrt{\cos(i)}$ arcsec) along the minor axis.  Three flattenings
of the bulge ($b/a=1,0.7,0.4$), four inclinations for the disk
($i=0^\circ,30^\circ,60^\circ,80^\circ$, where $i=0^\circ$ is
face-on and $i=90^\circ$ edge-on) and five values of the disk to bulge
ratio ($D/B=0,0.5,1,2,\infty$) are considered.  The resulting models
are normalized to $10^7$ counts. The pixel size is 0.6 arcsec. 
The circularly averaged luminosity profiles are derived following the 
same procedure adopted for the observed galaxies (see Paper III) and extend
out to $\approx 4-6R_e$. A 1\% sky error is introduced and 
the sky fitting procedure is activated. Note that the maximum
flattening of the EFAR galaxies is $b/a=0.5$, with 96\% of the
galaxies having $b/a>0.6$ (see Paper III). This corresponds to (pure) disk
inclinations $i\le 60^\circ$.

\begin{figure}
\plotone{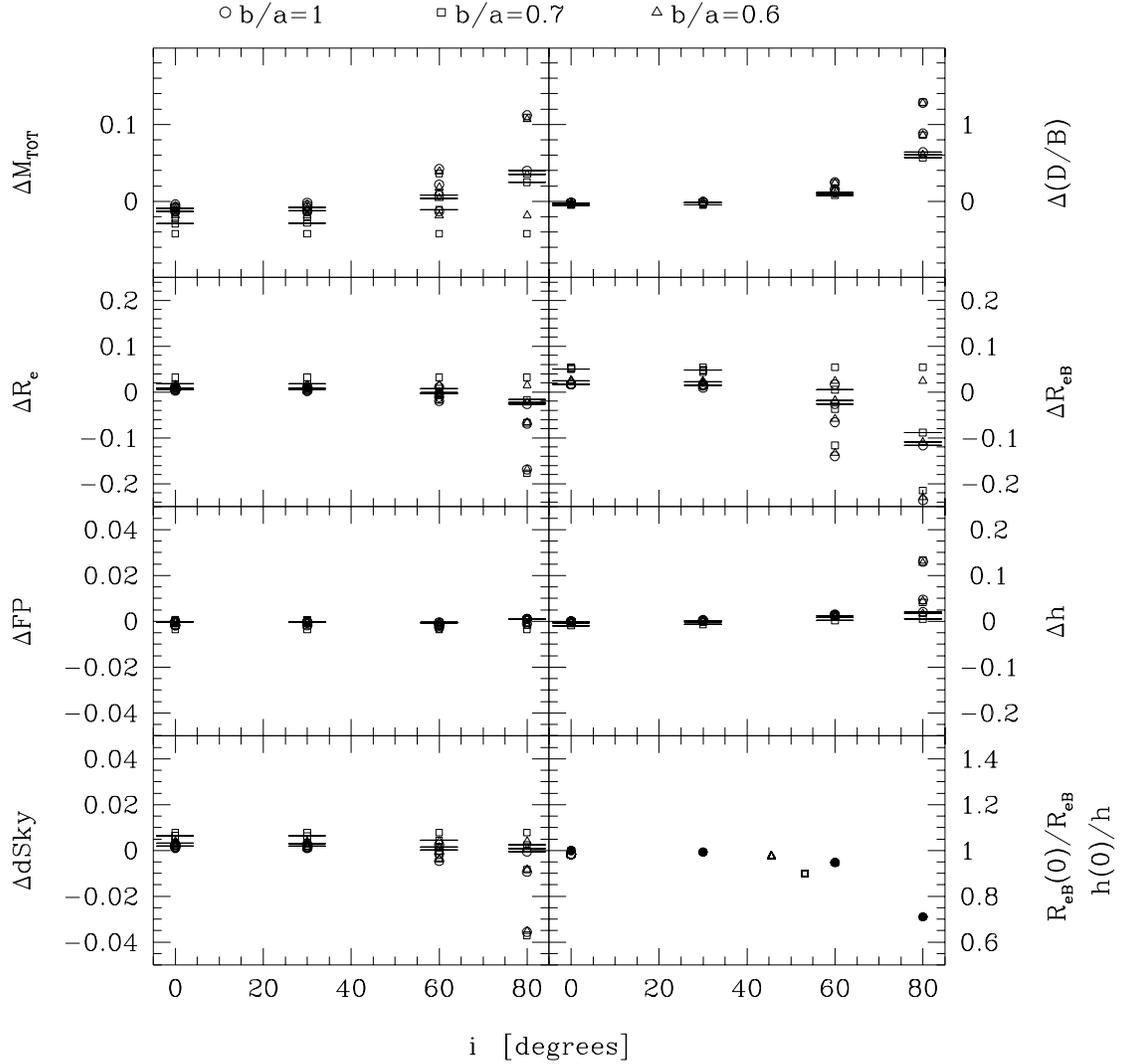}
\caption[f13.eps]{The reconstructed parameters of the bulge plus disk 
models as a function of the inclination $i$ of the disk.  Different symbols
indicate different flattenings of the bulge. The horizontal bars show
models with $D/B=0.5$. The plot at the bottom right
shows the scale lengths of the flattened bulge (open symbols) or of the 
inclined disk (filled symbols)
as a function of the flattening angle ($i=arccos(b/a)$) or of the
inclination angle, normalized to the $b/a=1$ or $i=0^\circ$ values.
Good reconstructions of the parameters are obtained when the
inclination is less than $60^\circ$ (see \S \ref{decomposition}).}
\label{figbulgedisk}
\end{figure}

Figure \ref{figbulgedisk} shows the reconstructed parameters as a
function of the inclination angle of the disk, for the different
flattenings of the bulge, using the sky fitting procedure. The
horizontal bars show models with $D/B=0.5$.
The plot at the bottom right shows the scale lengths of the flattened
bulge (filled symbols) or of the inclined disk as a function of the
flattening angle (open symbols, $i=arccos(b/a)$) or of the inclination
angle, normalized to the $b/a=1$ or $i=0^\circ$ values. When $D/B$ is
low ($\le 0.5$) the errors are very small for {\it every} inclination angle.
 For larger values of $D/B$, reliable
photometric parameters are obtained for $i<60^\circ$, but as soon as
the disk is nearly edge-on, total magnitudes and half-luminosity radii
are overestimated (by 0.1 mag and 20\% respectively). The integrated
circularized profiles, in fact, converge more slowly than the ones
following the isophotes. The correlated errors $\Delta FP$ always remain  
very small. Similarly, the parameters of the two
components are reconstructed well for $i<60^\circ$, but badly underestimate
the disk when it is nearly edge-on. However, a decent
fit is obtained, by increasing the half-luminosity radius of the bulge
component (see Fig. \ref{figbadfit} (e) and (f)). 
The sky correction is returned to better than
0.5\%  for $i<80^\circ$. The systematic effects
connected to the flattening of the bulge are small for the range of
ellipticities considered here ($b/a\ge0.6$).

These results indicate two potential problems, (i) galaxies
may be misclassified due to the presence of an edge-on disk component
not being recognized, or (ii) the photometric parameters may be 
systematically overestimated. However, these problems
do not apply to the EFAR sample, where $b/a>0.5$ always
and $b/a\ge 0.6$ for 96\% of the galaxies. Therefore, galaxies with bright
edge-on disks are only a very small fraction. Galaxies with faint edge-on
disks, which may not show large averaged flattenings, have low $D/B$
ratios and therefore are not affected by problem (ii). In a future
paper we will address the question whether in these cases the isophote
shape analysis might detect these faints disks and improve on point (i).

Finally, the two-dimensional frames described here have been used to calibrate
the estimator of the galaxy light contamination described in \S 
\ref{diskbulge}. We measured the sky in the same way as for the real
frames of Paper III,
by considering some small areas around the simulated galaxies. We find that
the predicted galaxy light contamination overestimates the measured sky 
excess by at least a factor two, and therefore can be used as a rather robust
upper limit to the galaxy light contamination.

%\placefigure{figbulgedisk}
%

\subsection{$R^{1/n}$ luminosity profiles}
\label{profiles}

The tests described above show that our fitting algorithm is able to
reconstruct the parameters of a sum of an
$R^{1/4}$ plus an exponential law accurately. In these cases sky subtraction
errors can also be corrected efficiently. Even so,  we do find in
Paper III that luminosity profiles of real early-type  
galaxies show systematic differences from $R^{1/4}$ plus exponential
profiles, yielding to a median reduced $\chi^2$ of 6. Here we quantify
the systematic effects that would be produced in this case, by
studying the case of the $R^{1/n}$ profiles.

CCO fitted the luminosity profiles of 52 early-type
galaxies using the  $R^{1/n}$ law introduced by Sersic (1968):
\begin{equation}
\label{r1n}
I(R)=I_e^n 10^{-b_n\left[\left(\frac{R}{R_e^n}\right )^{1/n}-1\right]},
\end{equation}
where $b_n\approx 0.868n-0.142$, $R_e^n$ is the half-luminosity
radius, and $I_e^n$ the surface brightness at $R_e^n$. 
The total luminosity is $L_T=K_nI_e^n{R_e^n}^2$, where $\log
K_n\approx 0.03[\log(n)]^2+0.441\log(n) +1.079$. Eq. \ref{r1n} reduces
to Eq. \ref{bulge} for $n=4$ and to Eq. \ref{disk} for $n=1$. 
For large values of $n$, Eq. \ref{r1n} describes a luminosity profile which 
is very peaked near the center and has a very shallow
decline in the outer parts.  Ciotti (1991) computes the curve of
growth related to Eq. \ref{r1n} analytically for integer values of
$n$ and finds that while already $\approx 13$\% of the total light is
included inside $R<0.05 R_e^n$, only 80\% of the total light is
included inside $6R_e^n$ for $n=10$.

\begin{figure}
\plotone{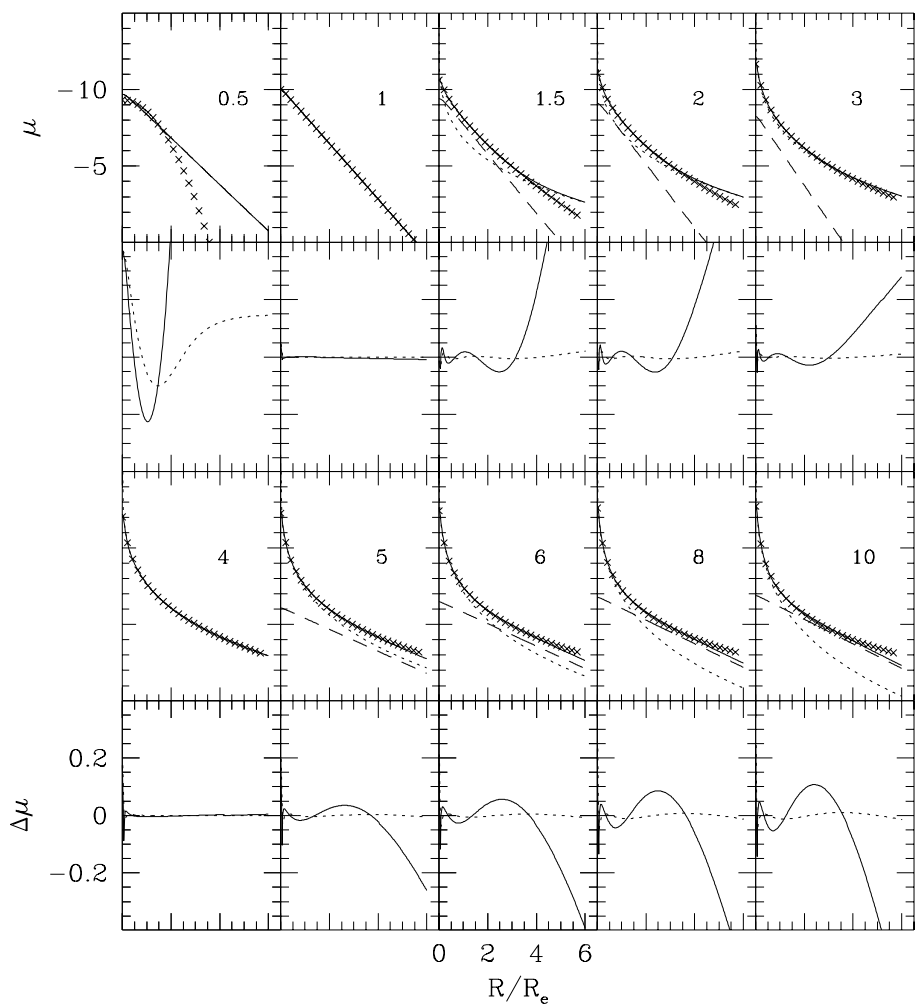}
\caption[f14.eps]{The fits to the $R^{1/n}$ law. Two plots are drawn 
for each value of $n$ (given in the top right corner). In the top plot the
crosses (one point in every seven) show the luminosity profiles
$\mu(R)=-2.5 \log I(R)$ of the
$R^{1/n}$ law as a function of $R/R_e$.  The dotted and dashed curves
show the best-fitting $R^{1/4}$ and exponential laws respectively. In
the bottom plot the residuals (full curves) in mag arcsec$^{-2}$ from the fits
to the $R^{1/n}$ law are shown. The dashed curve shows the residuals
(in mag) from the curves of growth (see discussion in \S \ref{profiles}).}
\label{figr1nfit}
\end{figure}

\begin{figure}
\plotone{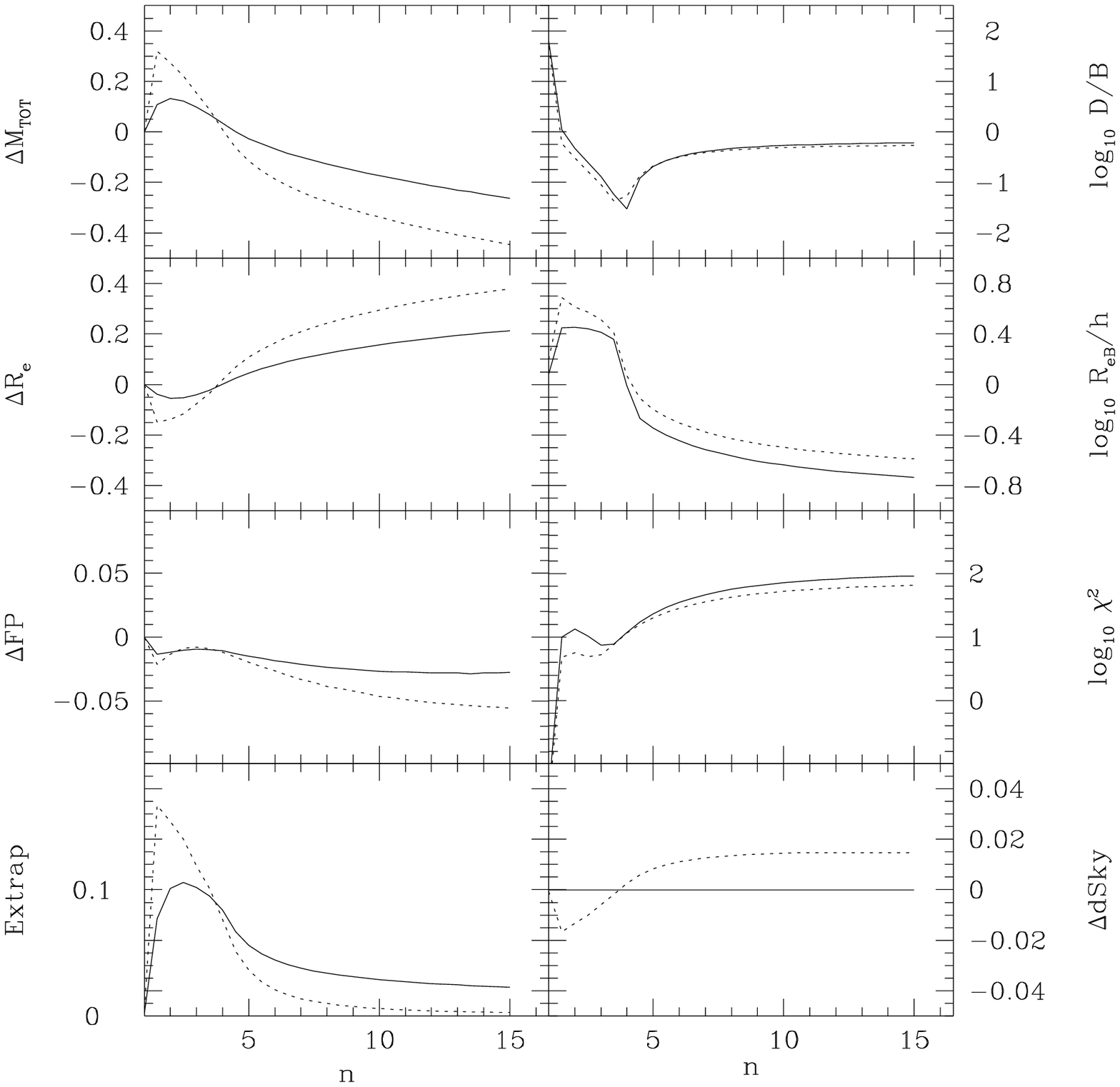}
\caption[f15.eps]{The relation between $n$ and the parameters of the 
decomposition (see \S \ref{profiles}). 
The full curves refer to the results obtained with no sky subtraction errors.
The dotted curves show the results obtained when the sky fitting algorithm 
is activated.}
\label{figr1npar}
\end{figure}

We fitted Eq. \ref{r1n}, 
modified to have a core at $R<0.05 R_e^n$, to an $R^{1/4}$ plus
exponential model for
$n=0.5$ to $n=15$ out to $6R_e^n$. Fig. \ref{figr1nfit} shows the
results of the fit for a selection of models.  With the
exception of the $n=0.5$ model, all of the $R^{1/n}$ profiles can be
described by a combination of an $R^{1/4}$ and an exponential
component, with residuals less than 0.2 mag arcsec$^{-2}$ for $R\le 4R_e$. 
For $n<4$ the 
residuals increase to 0.4 mag arcsec$^{-2}$ at $R>5R_e^n$, where the fits are
increasingly brighter than the $R^{1/n}$ profiles. For large
values of $n$ the residuals reach -0.4 mag arcsec$^{-2}$ at $R>5
R_e^n$, where the
fits are increasingly fainter than the $R^{1/n}$ profiles. The
relation between $n$ and the parameters of the decomposition is shown
in Fig. \ref{figr1npar}. Models with $1<n<4$ are fitted using a
decreasing amount of the exponential component, with a scale length
comparable to the one of the $R^{1/4}$ component. Models with $n>4$
are fitted with an increasing amount of the exponential component,
with increasingly large scale length.  Half-luminosity radii are
progressively underestimated, being $\approx 60$\% of the true values at
$n=15$. Correspondingly, total magnitudes are also underestimated, by
0.25 magnitudes at $n=15$.

A possible problem can emerge for large values of $n$, if
the sky fitting algorithm is activated. The dotted curves in
Fig. \ref{figr1npar} show that if the sky subtraction algorithm is
activated (Eq. \ref{chidmax}), then larger systematic effects are
produced.  Note that the computed sky correction (dotted curve of
Fig. \ref{figr1npar}) is $\approx 0$\ for $n \approx 1$ or $n\approx
4$ only. For
$n>4$ the correction is used to reduce the systematic negative
differences in the outer parts of the profiles.  A comparison between
the fitted sky corrections and the upper limits on the possible galaxy
light contamination (see \S  \ref{diskbulge} and
\ref{decomposition}) gives an important consistency check.  In the
case shown in Fig. \ref{figr1npar} the fitted sky corrections are
twice as large as the upper limits on the galaxy light
contamination. In a real case this, together with the rather large values of
$\chi^2$, would hint at an uncertain fitted sky correction.

The fact that the $R^{1/n}$ sequence can be approximated by a
subsample of $R^{1/4}$ plus exponential models suggests a possible
reinterpretation of CCO's results: the variety of profile shapes of
early-type galaxies is caused by the presence of a disk component.
Moreover, the use of the $R^{1/n}$ profiles to determine the
photometric parameters of galaxies of large $n$ is dangerous, since
the extrapolation involved is large and the fitted
profiles barely reach 2 or 3$R_e^n$, as derived from the fit. 
This problem is much smaller using the $R^{1/4}$ plus exponential
approach.

%\placefigure{figr1nfit}
%\placefigure{figr1npar}
%

\subsection{Discussion}
\label{discussion}

We conclude our tests by discussing the quality parameters defined in Table
\ref{tabquality} and their use to estimate the size of the systematic
errors present. 

The definitions given in Table \ref{tabquality} have been derived
after inspection of Figures \ref{figresdb} to \ref{figr1npar}, with
the desired goal of identifying three classes of precision, $\Delta M_{TOT}\le
0.05$, $\Delta M_{TOT}\le 0.15$, $\Delta M_{TOT}>0.15$. 
The parameters $Q_E$, $Q_{max}$,
$Q_{\chi^2}$, $Q_{S/N}$, and $Q_\Gamma$ are directly
related to the simulations.  Their low values imply  that
the fits involve a small extrapolation, extend to large enough radii,
give low surface brightness residuals with a large enough
signal-to-noise 
ratio and good spatial resolution. The definitions of $Q_{\hbox{Sky}}$
and $Q_{\delta \hbox{Sky}}$ deal with the accuracy of the sky
subtraction, taking into
account that high surface brightness galaxies suffer less from this
problem, and that large sky corrections indicate a lower quality of
the data. Low values of $Q$ (see Eq. \ref{qtot}) imply low values of all
quality parameters. 

\begin{figure}
\plotone{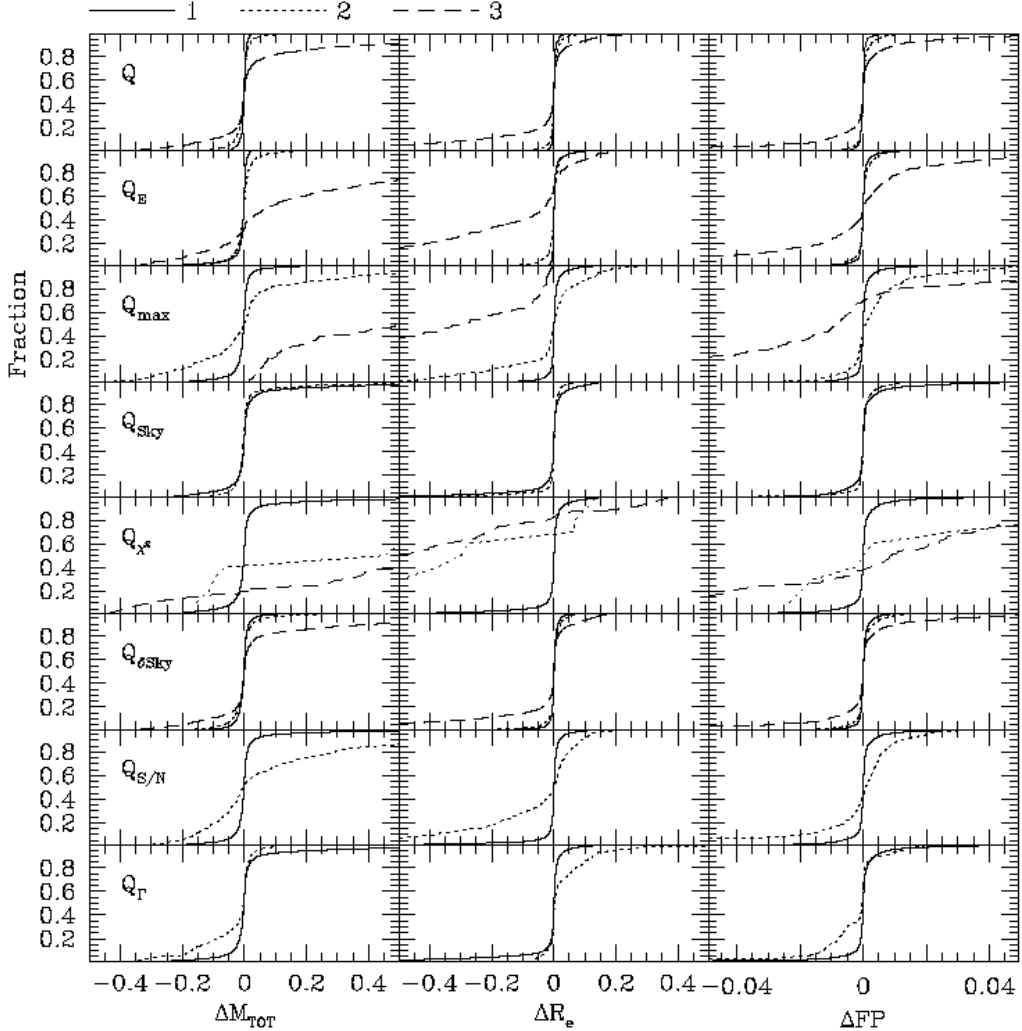}
\caption[f16.eps]{The precision of the reconstructed total 
magnitudes $M_{TOT}$,
the half-luminosity radii $R_e$ and the combined quantity $FP=\log R_e - 0.3
\langle SB_e\rangle $.
The cumulative distributions of the errors
$\Delta M_{TOT}$, $\Delta R_e$ and $\Delta FP$ derived from  
all the performed disk plus bulge fits with sky correction algorithm 
activated are shown as a function of the different quality parameters 
defined in \S \ref{quality}. The full lines plot the distributions when
the parameters have value of 1, the dotted ones when the value is 2, 
the dashed ones when the value is 3. 
The distributions derived by selecting on the global
quality parameter $Q$ match the precision ranges identified in \S 
\ref{discussion}.}
\label{figqualdis}
\end{figure}

Figure \ref{figqualdis} shows the cumulative distributions of the errors
$\Delta M_{TOT}$, $\Delta R_e$ and $\Delta FP$ derived from  
all the performed disk plus bulge fits with sky correction algorithm 
activated, as a function of the different quality parameters. The two
most important parameters regulating the precision of the photometric 
parameters are the level of extrapolation and the goodness of the fit, followed
by the sky subtraction errors. A low $Q_E$ fixes the maximum possible 
overestimate of the parameters. A low $Q_{\chi^2}$ with a low $Q_E$ 
constrains the underestimate and the reliability of the sky correction. 
The ranges of the errors match the desired goal of identify three classes 
of precisions.

Finally, it is sobering to note that the constraints
needed to achieve $Q=1$, high precision total magnitudes and $R_e$ are
rather stringent.  Only 16\% of EFAR galaxies have $Q=1$. 
Most of the existing {\it published} values of $M_{TOT}$ and $R_e$
of galaxies are far below this precision, because of the restricted
radial range probed by photoelectric measurements or small CCD chips, 
because of sky subtraction errors, and also by the use of 
the pure $R^{1/4}$ curve of growth fitting (see Figure \ref{figquatexp}). 
The related observational problems
can be somewhat reduced with the use of large CCDs (see Introduction), 
but the {\it a priori} limiting factor of galaxy 
photometry, the extrapolation, will always remain with us at a certain level. 

On the other hand, the errors on $M_{TOT}$ and $R_e$ are strongly
correlated, so that the quantity $\log R_e-0.3\langle SB_e \rangle$ is
always well determined. This fact allows the
accurate distance determinations achieved using the Fundamental Plane
correlations despite the systematic errors in the photometric quantities.

%\placefigure{figqualdis}
%

\section{Conclusions}
\label{conclusions}

We constructed an algorithm to fit the circularized profiles of the
(early-type) galaxies of the EFAR project, using a sum of a
seeing-convolved $R^{1/4}$ and an exponential law. This choice allows
us to fit the large variety of profiles exhibited by the
EFAR galaxies homogeneously. The procedure provides
for an optimal combination of multiple profiles. A sky fitting option 
has been developed. A conservative upper limit to the sky
contamination due to the light
of the outer parts of the galaxies is estimated.
From the tests described in previous sections we draw the following 
conclusions:

1) The reconstruction algorithm applied to simulated $R^{1/4}$ plus
exponential profiles shows that random errors are negligible if the
total signal-to-noise ratio of the profiles exceeds $300$.  Systematic
errors due to the radial extent of the profiles are minimal if
$R_{max}/R_e> 2$.  Systematic errors due to sky subtraction are
significant (easily larger than 0.2 mag in the total magnitude) 
when the sky surface brightness is of the order of the
average effective surface brightness of the galaxy. They can be
reliably corrected for as long as the fitted profiles show small
systematic deviations ($\chi^2<12.5$).

2) Strong systematic biases (errors larger than 0.2 mag in the total
magnitudes) are present when a simple $R^{1/4}$ or exponential
model is used to fit test profiles with disk to bulge ratios as low as 0.2.

3) The use of the shape of the (normalized) $\chi^2$ function badly
underestimates the (systematic) errors on the photometric parameters.

4) Systematic biases emerge when test profiles are derived for systems with
significant disk components seen nearly edge-on, or when the fitted luminosity
profile declines more slowly than  an $R^{1/4}$ law. The parameters
of bulge plus disk systems can be determined to better than $\approx 20$\%
if the disk is not very inclined ($i<60^\circ$). 
 
5) The sequence of $R^{1/n}$ profiles, recently used to fit the
profiles of elliptical galaxies by Caon et al. (1993), is equivalent to
a subset of $R^{1/4}$ and exponential profiles, with appropriate
scale lengths and disk-to-bulge ratios, with moderate systematic biases
for $n\le 8$ and residuals less than 0.2 mag arcsec$^{-2}$ for $R\le
4R_e$. This suggests that the variety
of luminosity profiles shown by early-type galaxies is due to the
frequent presence of a weak disk component.

6) A set of quality parameters has been defined to control the
precision of the estimated photometric parameters. They take into
account the amount of extrapolation involved to derive the total
magnitudes, the size of the sky correction, the average surface
brightness of the galaxy relative to the sky, the radial extent of
the profile, its signal-to-noise ratio, the seeing value and the
reduced $\chi^2$ of the fit. These are combined into a single quality
parameter $Q$ which correlates with the expected precision of the
fits.  Errors in total magnitudes $M_{TOT}$ less than 0.05 mag and in
half-luminosity radii $R_e$ less than 10\% are expected if $Q=1$, and
less than 0.15 mag and 25\% if $Q=2$.

89\% of the EFAR galaxies have
fits with $Q=1$ or $Q=2$. The errors on the combined Fundamental Plane
quantity $FP=\log R_e -0.3\langle SB_e\rangle$, where $\langle SB_e
\rangle$ is the average effective surface brightness, are smaller than
0.03 even if $Q=3$. Thus systematic errors on $M_{TOT}$ and $R_e$
 only marginally  affect the distance estimates which involve $FP$.

\acknowledgments {RPS acknowledges the support by DFG grants SFB 318
and SFB 375. GW is grateful to the SERC and Wadham College for a
year's stay in Oxford, and to the Alexander von Humboldt-Stiftung for
making possible a visit to the Ruhr-Universit\"at in Bochum. MMC
acknowledges the support of a Lindemann Fellowship, a DIST Collaborative
Research Grant and an Australian Academy of Science/Royal Society
Exchange Program Fellowship. This work was partially supported by NSF
Grant AST90-16930 to DB, AST90-17048 and AST93-47714 to GW,
AST90-20864 to RKM, and NASA grant NAG5-2816 to EB. The entire
collaboration benefitted from NATO Collaborative Research Grant 900159
and from the hospitality and monetary support of Dartmouth College,
Oxford University, the University of Durham and Arizona State
University. Support was also received from PPARC visitors grants to
Oxford and Durham Universities and a PPARC rolling grant:
``Extragalactic Astronomy and Cosmology in Durham 1994-98''.}

\appendix
\section{The fitting function}
\label{fittingfunction}

The fitting procedure described in \S \ref{fitting} assumes that fitted
profiles can be well represented by the sum of a de Vaucouleurs (1948)
law of
half-luminosity radius $R_{eB}$ and a surface brightness $I_{eB}$ at $R_{eB}$ 
(with $B$ for {\it bulge} component):
\begin{equation}
\label{bulge}
I_B(R)=I_{eB} \exp\{-7.67[(R/R_{eB})^{1/4}-1]\},
\end{equation}
and an exponential component with
exponential scale-length $h$ and central surface brightness $I_0$ 
(with $D$ for {\it disk} component):
\begin{equation}
\label{disk}
I_D(R)=I_0 \exp(-R/h).
\end{equation}

The $R^{1/4}$ law curve-of-growth is:
\begin{equation}
\label{bulgegrowth}
F_B(R)=L_B\left[1-\exp(-z)\left(1+\sum_{n=1}^7\frac{z^n}{n!}\right)\right],
\end{equation}
where the total luminosity of the bulge component is normalized,
$L_B=7.22\pi I_{eB} R_{eB}^2=1$, and $z=7.67 (R/R_{eB})^{1/4}$. The
exponential law curve-of-growth is:
\begin{equation}
\label{diskgrowth}
F_D(R)=L_D[1-(1+R/h)\exp(-R/h)],
\end{equation}
where the total luminosity of the disk component is set to the disk-to-bulge
ratio, $L_D=2\pi I_0h^2=(D/B)$, if a two-component model is
considered, or normalized, $L_D=1$, if an exponential only model is
used (in this case $L_B=0$).

Both laws are seeing convolved with a $\gamma=1.6$ psf, 
following the technique described by Saglia et al. (1993). The Fourier 
transforms of the $\gamma$ psfs are given by:
\begin{equation}
\label{gammapsf}
\hat{p}_\gamma(k) \equiv \int_0^\infty \, 2\pi R \, J_0(kR) \, p_\gamma(R) \,
dR \, = \exp[-(kb)^\gamma],
\end{equation}
where $J_0(kR)=\frac{1}{2\pi}\int_0^{2\pi}\exp (ikR\cos \theta)d\theta$ is 
the zero-order Bessel function.
The $\gamma=1.6$ psf reproduces well the stellar profiles measured
with the telescopes and setups used in Paper III (see Saglia et al. 1993).

A grid of seeing-convolved models is obtained for 100 values of the
$\Gamma/R_{eB}$ and $\Gamma/h$ ratios, ranging from 0.01 to 1 with 
linear increment of 0.01. Here $\Gamma$ is the FWHM of
the seeing profile. For each of these values, the seeing-convolved luminosity
profiles $I^C_B(R/\Gamma, \Gamma/R_{eB})$ and
$I^C_D(R/\Gamma,\Gamma/h)$ and curves of growth
$F^C_B(R/\Gamma,\Gamma/R_{eB})$ and $F^C_D(R/\Gamma,\Gamma/h)$ for
both the bulge and the disk component are tabulated for
$0<R/\Gamma<50$ on a logarithmic radial grid (plus $R=0$) with $d\ln
R/\Gamma=0.230258$ and 31 points.  A cubic spline interpolation on
$\ln(R/\Gamma$) and a linear interpolation on $\Gamma/R_{eB}$ or $
\Gamma/h$ are used to determine the profile at a given radial distance
$R$ and with given values for $R_{eB}$, $h$ and $\Gamma$. When
$R/\Gamma<0.05$ a log-log extrapolation is used. When
$R/\Gamma>50$ the correction computed for $R/\Gamma$=50 is applied.
If $\Gamma/R_{eB}<0.01$ ($\Gamma/h<0.01$) the correction computed for
$\Gamma/R_{eB}=0.01$ ($\Gamma/h=0.01$) is used. When $\Gamma/R_{eB}>1$
($\Gamma/h>1$) the correction computed for $\Gamma/R_{eB}=1$
($\Gamma/h=1$) is applied. The resulting numerical errors in the
seeing convolved bulge and disk models are negligible ($<<1$\%).

The luminosity profile $f_{B+D}=f_B+f_D$ fitted to the data takes into account
 the effect of the finite pixel size of the observed profiles. 
These are computed as the azimuthally averaged flux in the annulus of
radius $R$ and of half pixel width. This procedure is reproduced by the
following equations:
\begin{equation}
\label{pixelbulge}
f_B(R,R_{eB},\Gamma,S)=\frac{1}{A}\left\{F^C_B\left[\frac{R+S/2}\Gamma,
\frac{\Gamma}{R_{eB}}\right]-F^C_B\left[\frac{R-S/2}\Gamma,\frac
{\Gamma}
{R_{eB}}\right]\right\},
\end{equation}
\begin{equation}
\label{pixeldisk}
f_D(R,h,\Gamma,S)=\frac{1}{A}\left\{F^C_D\left[\frac{R+S/2}\Gamma,
\frac{\Gamma}{h}\right]-
F^C_D\left[\frac{R-S/2}\Gamma,\frac{\Gamma}{h}\right]\right\},
\end{equation}
where $A=\pi[(R+S/2)^2-(R-S/2)^2]$ is the area of the annulus and $S$
is the scale or pixel size in arcsec. Eqs. \ref{pixelbulge} and
\ref{pixeldisk} are valid for $R>S/2$. If $R<S/2$ (i.e., the central
value at $R=0$), then:
\begin{equation}
\label{pixelbulge0}
f_B(R,R_{eB},\Gamma,S)=\frac{ F^C_B\left[
\frac{R+S/2}{\Gamma},\frac{\Gamma}{R_{eB}}\right]}
{\pi (R+S/2)^2}
\end{equation}
\begin{equation}
\label{pixeldisk0}
f_D(R,h,\Gamma,S)=\frac{ F^C_D\left[
\frac{R+S/2}{\Gamma},\frac{\Gamma}{h}\right]}{\pi (R+S/2)^2}.
\end{equation}

Similar tables of seeing convolved profiles were also constructed for the
$\Psi=12$ $f_\infty$ model and for the smoothed $R^{1/4}$ used in Saglia et 
al. (1993). The luminosity profile of the $\Psi=12$ $f_\infty$ 
model is more centrally peaked than the $R^{1/4}$ and 
declines less rapidly than the $R^{1/4}$ law at large radii. The smoothed
$R^{1/4}$ model is less centrally concentrated than the $R^{1/4}$ law.
Both profiles have been used to test our fitting algorithm (see \S 
\ref{seeing}).

\end{document}